\documentclass[pra, twocolumn, final, reprint, amsmath, amssymb, superscriptaddress, showpacs, a4paper, aps, 10pt, hidelinks, numbers, sort&compress, floats, balancelastpage]{revtex4-1}

\usepackage{graphicx, color} 
\usepackage[normalem]{ulem}          

\usepackage[colorlinks=true]{hyperref}
\hypersetup{pdfpagemode=None, linkcolor=black, citecolor=blue, urlcolor=blue}  

\usepackage[utf8]{inputenc}

\usepackage[caption=false]{subfig}
\usepackage{amssymb}
\usepackage{amsmath}
\usepackage{commath}
\usepackage{graphicx,bm}
\usepackage{verbatim}

\definecolor{britishracinggreen}{rgb}{0.0, 0.26, 0.15}
\definecolor{bulgarianrose}{rgb}{0.28, 0.02, 0.03}
\definecolor{darkred}{rgb}{0.90,0,0}
\definecolor{darkgreen}{rgb}{0,0.60,.2}
\definecolor{darkblue}{rgb}{0,0,1}
\definecolor{orange}{cmyk}{0,0.6,0.8,0}
\definecolor{lightblue}{rgb}{0.3,0.5,1}
\definecolor{lightgreen}{rgb}{0.4,0.80,.4}

\begin{document}

\title{Dissipation in a Finite Temperature Atomic Josephson Junction}

\author{K. Xhani$^{1,2}$ and N. P. Proukakis}

\address{
Joint Quantum Centre (JQC) Durham-Newcastle, School of Mathematics, Statistics and Physics, \\Newcastle University, Newcastle upon Tyne NE1 7RU, United Kingdom\\
\mbox{$^{2}$ CNR-INO, European Laboratory for Non-Linear Spectroscopy (LENS), 50019 Sesto Fiorentino, Italy}}

\date{\today}



\begin{abstract}


We numerically demonstrate and characterize the emergence of distinct dynamical regimes of a finite temperature bosonic superfluid in an elongated Josephson junction generated by a thin Gaussian barrier
over the entire temperature range where a well-formed condensate can be clearly identified.
%
Although the dissipation arising from the coupling of the superfluid to the dynamical thermal cloud increases with increasing temperature as expected, the importance of this mechanism  is found to depend on two physical parameters associated (i) with the initial chemical potential difference, compared to some characteristic value, and (ii) the ratio of the thermal energy to the barrier amplitude.
The former determines whether the superfluid Josephson dynamics are dominated by gradually damped plasma-like oscillations (for relatively small initial population imbalances), or whether dissipation at early times is instead dominated by vortex- and sound-induced dissipation (for larger initial imbalances).
%
The latter defines the effect of the thermal cloud on the condensate dynamics, with a reversal of roles, i.e.~the condensate being driven by the oscillating thermal cloud, being observed when the thermal particles acquire enough energy to overcome the barrier.
Our findings are within current experimental reach in ultracold superfluid junctions.

\end{abstract}

\maketitle

\section{Introduction}

Josephson effects across a junction separating two parts of a quantum liquid~\cite{JOS,anderson63} are ubiquitous in nature, occurring across superconductors ~\cite{JOS,anderson63,barone}, superfluid helium ~\cite{avenel,davis-helium,Sato19,hoskinson-2006} and trapped ultracold atomic gases~\cite{Anderson-Kasevich,JO0,JO1,Anker05,Schumm05,Shin05,Steinhauer07,JO5,LeBlanc11,Liscience,Lidiss,Kwon84,Luick20,delpace2020tunneling}, and exciton-polariton condensates ~\cite{Lagoudakis10,Adiyatullin2017}.
%
The most characteristic manifestation of Josephson junctions in ultracold atomic systems relates to the so-called `plasma' oscillations associated with periodically alternating particle transfer across the junction with a population difference and an associated relative phase between the two sides of the junction  oscillating about a zero value. Such behaviour has been observed in a range of experiments, 
including ultracold bosonic atoms in diverse geometries~\cite{JO0,JO1,Schumm05,Steinhauer07,LeBlanc11,JO5}
fermionic superfluids across the BEC-BCS crossover \cite{Liscience,Lidiss,Kwon84,delpace2020tunneling,Luick20}, with atomic current across a Josephson junction playing an important role in atomtronics \cite{Edwards2013,ryu_13,ryu_20,atomtronic,gauthier_19}.

As the underlying features are of a quantum nature, associated with the densities and phases of the superfluids across the two sides of the junction, most theoretical treatments to date have focussed on the more fundamental, pure superfluid (zero-temperature) analysis. 
The dynamics of such atomic systems was first analysed in Ref.~\cite{MQST1,MQST3}, who predicted two distinct regimes, with their analysis based on a two-mode model and a corresponding analogy to the non-rigid pendulum  with variable length with more detailed theoretical/numerical analysis conducted by various authors \cite{MQST2,Sakellari_2004,MQST2,Abad,Pascucci20,spuntarelli2007}.
Specifically, the work of Ref.\cite{MQST1,MQST3} highlighted the existence of a `self-trapping' regime characterized by population imbalance oscillations about a non-zero value (i.e.~one side of the junctions always maintain higher population than the other), associated with a relative phase increasing  in time (rather than an oscillating) phase. Such a regime, was first  observed in \cite{JO1}, followed by related experiments in other groups \cite{Steinhauer07,JO5}.

Viewed differently, the Josephson junction separating two parts of a superfluid can also be viewed as a barrier acting against the underlying superflow: as such, superflow dissipation can emerge even strictly at $T=0$, by the generation of sound waves, and even nonlinear excitations, such as solitons \cite{polo_19,Saha21}, vortices \cite{Xhani20,XhaniNJP,Liscience,Lidiss,Jendrzejewski,Eckel,Piazza1} and shock waves \cite{Saha21,griffin2019},  depending on system geometry and dimensionality. Such dynamical excitation features across a Josephson junction, well-known as phase slips in the context of superconductors \cite{halperin,Chen2014}, superfluids~\cite{avenel,hoskinson-2006} ultracold transport \cite{Jendrzejewski,Eckel,Xhani20,polo_19,Wright2013,Abad,gauthier_19,Yakimenko-PhaseSlips}, have also been observed in recent ultracold experiments with fermionic superfluids \cite{Liscience,Lidiss}.
Such behaviour has been previously analysed in depth by the present authors \cite{Xhani20}, thus shedding more light into the microscopic and energetic origins of dissipation in such systems, and directly connecting microscopic findings with experimental observations both for coherent and dissipative transport across an ultracold Josephson junction.

The combination of earlier experimental and theoretical works thus indicated two distinct transitions from the Josephson `plasma' regime with increasing initial population imbalance, namely to either the self-trapped or to the phase-slip-induced dissipative dynamical regimes.
The former, `self-trapped', regime is generally expected to emerge in the limit of validity of the two-mode model~\cite{MQST1,MQST3,Smerzi03,andrea2003,ragh,Marino,Kylstra,Sakellari_2004,Zou-Dalfovo}, i.e.~ in the limit of rather high barriers, compared to the system chemical potential.
To better understand the conditions leading to either of these two dynamical transitions, the present authors undertook a detailed systematic analysis of the phase diagram of the dynamical regimes across a Josephson junction as a function of initial population imbalance, barrier properties 
and underlying system geometry
\cite{XhaniNJP}. Our findings clearly characterized the nature of the transition of the Josephson oscillations with increasing population imbalance as being `dissipative' in the limit of relatively low/narrow barriers, and `self-trapped' in the opposite regime of high/wide barriers, with a complicated intermediate regime featuring irregular population/phase dynamics, thus providing a complete characterization of the emergence of such different dynamical regimes. 
Such analysis was motivated by the experimental set up at LENS~\cite{Lidiss, Liscience}, in which a molecular BEC of fermionic $^{6}Li$ atoms was placed in an elongated 3D geometry featuring a narrow Gaussian barrier along its axial direction.
Specifically for the experimental parameters
%
in which  $w/\xi \sim 4$ (where $w$ is the width of the Gaussian barrier and $\xi$ the superfluid healing length), our numerical analysis demonstrated the appearance of the dissipative regime for $V_0/\mu \lesssim 1.2$,  whereas 
self-trapping only emerged in a clear manner for $V_0/\mu \gtrsim 1.6$ (or when considering higher values of $w/\xi$ than probed experimentally) -- see Ref.~\cite{XhaniNJP} for more details.

While such work thoroughly addressed the $T=0$ pure superfluid dynamical regimes, few studies to date have studied the effects of thermal, or quantum,  excitations \cite{MQST4,Ic2,MQST2,piselli20,ragh,gati2006,quantum_fluc,Franzosi,milburn1997,singh20,polo_19,polo2018} . 
This can be crucial, since 
experiments are typically performed at small, but non-zero, temperatures $T \ll T_{c}$ (where $T_c$ is the critical temperature for Bose-Einstein condensation): in fact, evidence of thermal dissipation in the self-trapped regime has already been experimentally observed in \cite{Steinhauer07}. 
It is thus of significant interest to understand the role of the thermal cloud on the junction dynamics. Some work on the decay of the self-trapped regime has been performed in the context of the dissipative and stochastic projected Gross-Pitaevskii equation \cite{MQST4}, qualitatively reproducing the findings of \cite{Steinhauer07}.


In the present work, we provide a unified characterization of the fundamental role of thermal dissipation in both the Josephson  plasma and the dissipative regimes, by means of a self-consistent theory which incorporates a dynamical thermal cloud and its back action on the condensate.
Specifically, we perform a detailed analysis of the long-term dissipative dynamical evolution of the superfluid across a Josephson junction, focussing on the relative population dynamics, their dominant frequencies, and the relation between condensate and thermal cloud dynamics. 
We identify two distinct dynamical regimes, namely a low temperature regime in which the small thermal cloud is driven by the condensate, and a high temperature regime in which the thermal cloud has enough energy to overcome the barrier, and thus begins to drive the condensate.
The dominant frequencies identified are the Josephson plasma frequency (slightly lower than the trap frequency), a frequency we interpret as its corresponding second order contributions, 
and the dipolar frequency of the thermal cloud (which is close to the underlying harmonic trap frequency).
All such frequencies are found to be relevant in both the Josephson plasma and the dissipative regimes, with their relative importance dependent both on dynamical regime and on temperature, as we shall discuss.

This paper is structured as follows:
Sec.~\ref{sec2} introduces the key concepts required for our analysis, namely the physical system and geometry (Sec.~\ref{sec2} A), a brief summary of the underlying dynamical regimes in the limit of a pure superfluid (Sec.~\ref{sec2} B),
and of the documented importance of second-order tunneling contributions to date 
(Sec.~\ref{sec2} C), 
with the dynamical finite temperature model used summarized in Sec.~\ref{sec2} D. 
Sec.~\ref{sec3} summarizes the parameter regime of this study and identifies the physical observables which are used to analyze the emerging dynamics across the entire temperature domain (Sec.~\ref{sec3} A), analyzing the system dynamics, dominant frequency components and damping rates in both the Josephson 
(Sec.~\ref{sec3} B)
and dissipative regimes (Sec.~\ref{sec3} C), at fixed condensate number, further highlighting the role of the thermal cloud on the damping of sound waves.
Our findings on the finite temperature dynamical regimes are revisited in Sec.~\ref{sec4} in the context of fixed total particle number, 
with our observations further discussed and concluded in Secs.~\ref{sec5a} and \ref{sec5} respectively.


\section{Physical System and \\ $T>0$ Kinetic Model \label{sec2}}

\subsection{A Gaussian Junction in an Elongated Anisotropic Harmonic Trap}

The physical system considered in this study is an anisotropic highly elongated harmonic trap, with a double-well potential of the form
\begin{eqnarray}
V_{ext}(x,y,z)&=&\frac{1}{2}M \left( {\omega_x}^2 x^2+ {\omega_y}^2 y^2+ {\omega_z}^2 z^2 \right)
\nonumber \\
&+& V_0 \, e^{-2x^2/w^2}\;. \label{Vtrap}
\end{eqnarray}

This geometry is motivated by the LENS fermionic superfluid experiments which observed the dissipative regime, with our analysis restricted to  the BEC limit of lithium molecules \cite{Liscience,Lidiss}.
As such, and consistent with our earlier works \cite{Xhani20,XhaniNJP} our study uses
trapping frequencies 
$\omega_x=2 \pi \times 15$ Hz, $\omega_y=2 \pi \times 148$Hz, and $\omega_z=2 \pi \times 187.5$Hz
across the $x$, $y$ and $z$ directions respectively.
In order to avoid a potential change in the system dynamical regime caused by a changing condensate particle number with varying temperature, 
our primary finite temperature study is conducted at fixed condensate particle number, $N^{\rm BEC} = (5.04 \pm 0.02) \times 10^4$.
%
We also keep fixed the barrier height $V_0 = 104 \hbar \omega_x $, and width $w = 2\mu$m.
In the pure superfluid limit ($T=0$), this amounts to $V_0 =0.97 \mu$, and $w =3.8 \xi$, where $\xi=\hbar / \sqrt{2 \mu M}=0.52 \mu$m denotes the condensate healing length. 
As previously shown~\cite{Xhani20,XhaniNJP}, 
in such a regime, the system is far away from the deep tunneling regime and thus cannot support the emergence of macroscopic quantum self-trapping;
in fact, the two-mode model does not accurately predict the system behaviour for the chosen barrier height.

A schematic of the initial system density for such parameters is shown in Fig.~1 for $T=0$ [pure condensate, Fig.~1(a)] and $T=0.58T_c$ [Fig.~1(b)]: in the latter case, the top panel (i) shows the condensate, with the corresponding thermal cloud contribution shown -- within the context of the Hartree-Fock approximation -- in panels (ii), clearly revealing the thermal cloud surrounding the condensate and partly infilling the barrier region where the condensate contribution decreases. More details of our theoretical model are discussed in Sec.~II D and Appendix A.
%
%
The dependence of the condensate fraction on (scaled) temperature 
over which such dynamical behaviour is characterised is shown in Fig.~1(c).
As our analysis is performed with a fixed {\em condensate} particle number at different temperatures, this implies that the thermal, and thus total, particle number $N=N(T)$  increases with increasing temperature; for this reason, each point on this graph has been scaled by its own corresponding non-interacting harmonically trapped 3D critical temperature $T_c =  T_c(N(T)) \simeq 0.94 \times (\hbar \bar{\omega}/k_B)   (N(T))^{1/3}$ [where $\bar{\omega} = (\omega_x \omega_y \omega_z)^{1/3}$] \cite{RevModPhys.71.463}. 
For comparison, we also plot the non-interacting prediction for the condensate fraction, which confirms the expected shift of the critical region due to finite-size and mean-field corrections \cite{RevModPhys.71.463}.
As evident from Fig.~1(c) our analysis is performed over a very broad temperature range, in which there is at least a $10\%$ condensate fraction, thus avoiding limitations of our approach as the system approaches the critical region.

%


To seed the dynamics, we create an initial population imbalance by applying a linear barrier shift $-\epsilon x$ to the double-well potential, 
to fix the initial condensate imbalance for all $T$ considered
(a subtle shift at rather high $T$ will be commented upon later, in Sec.~III and Appendix E).

\subsection{Josephson Plasma Oscillations vs.~Dissipative Dynamics at $T=0$}

As well-known, a pure ($T=0$) superfluid system with an initial population imbalance across the junction can exhibit undamped oscillatory particle transfer across the junction (for a given barrier height/width configuration) with plasma frequency $\nu _J$ (shown in Fig.~1(d) by the black dashed line).
This is typically characterized by the fractional population imbalance
\begin{equation}
z(t)=\frac{N_R (t)-N_L (t)}{N_R (t)+N_L (t)} 
\end{equation}
where $N_{R/L}$ is the number of condensate particles on the right/left sides of the barrier (centered at $x=0$); the time derivative of this quantity gives the superfluid current across the junction, via the expression~$I=-(N/2) dz(t)/dt$, with $N=N_R+N_L$ being the total particle number.

\begin{figure}[h!]
\begin{center}
\includegraphics[width= \columnwidth]{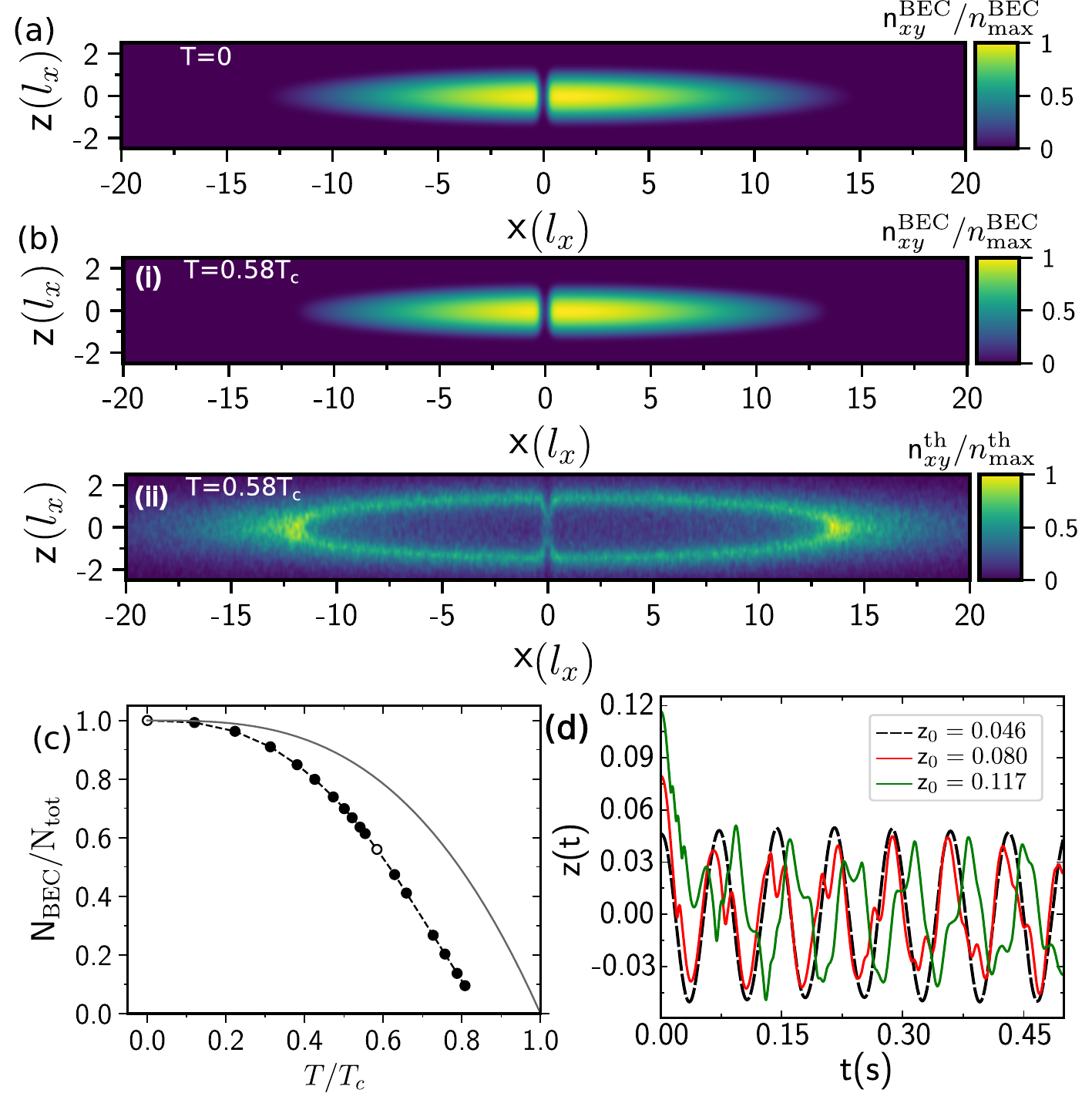}
\caption{
(a) Equilibrium 2D integrated condensate density in the $xz$-plane at $T=0$ (which coincides with the total density). (b) The corresponding equilibrium  integrated (i) condensate and (ii) thermal cloud density profiles in the $xz$-plane  at $T=$88nK $= 0.58 T_c$ for the same fixed condensate number $N_{\rm BEC} \simeq 5.04 \times 10^4$, and $V_0/\mu(T=0)=0.97$.
All densities are scaled to their corresponding maximum values, as indicated by the colourbars.
(c) Condensate fraction as a function of the temperature $T$ scaled to the non-interacting critical value $T_c=T_c(N(T))$ for the corresponding total particle number $N(T)$ in each case (circles), showing the entire regime of temperatures probed in our present analysis: the cases $T=0$ and $T=0.58T_c$ whose corresponding densities were shown respectively in (a)-(b) are indicated  as hollow circles, and the dashed line is a guide to the eye. The thin solid grey line shows the corresponding ideal gas analytical prediction.
(d) Time evolution of the population imbalance at $T=0$ for different values of $z_0$: (i) for an initial imbalance $z_0$ just below $z_{\rm cr}$ the system exhibits undamped Josephson plasma oscillations (black dashed line); (ii) at $z=z_{cr}$, corresponding to the critical imbalance for the onset of the dissipative regime, with a single vortex ring  generated (red line); (iii) $z_0 > z_{cr}$, showing kinks characteristic of the sequential generation of multiple vortex rings (green line). 
}  
\label{cond_therm_dens_zcr}
\end{center}
\end{figure}



Increasing the initial population imbalance for our parameters actually leads to flow speeds exceeding the local critical velocity, thus inducing the  dissipation of superflow kinetic energy through the generation of vortex rings and associated sound waves, an effect already interpreted in our previous work \cite{Xhani20,XhaniNJP}, and observed experimentally \cite{Lidiss,Liscience}.
This is the so-called dissipative regime, characterized by a rapid initial decay of
the population imbalance (during which vortex rings are nucleated), followed by oscillations around a zero value with one, or more, frequencies.
%
The emergence of such behaviour can be clearly seen in the evolution of the population imbalance corresponding to the red and green curves in Fig.~1(d).


As a first step, we explicitly calculate for our parameters the critical population imbalance marking the transition from Josephson plasma to phase-slip-induced dissipative regime (shown in Appendix C). 
 At $T=0$ and for each barrier height value, we define $z_{\rm cr} ^{\rm BEC}$ 
as the first explored value of initial imbalance at which the rapidly decaying population imbalance amplitude exhibits (within $t \sim 0.1$s) exactly one kink, while simultaneously  the relative phase undergoes one phase-slippage  mechanism: 
such behaviour is shown by the red profile in Fig.~1(d), 
 with the single kink around 20ms indicating the backflow associated with the generation of a single vortex ring (and associated acoustic emission) \cite{Xhani20, XhaniNJP}. 
 Further increase in the population imbalance leads to sequential generation of multiple vortex rings across the junction during the early phase of uni-directional flow across the junction, as can be seen by the green line showing the case of an initial population imbalance well above the critical value for the chosen parameters.
We use the same criterion to identify the critical condensate population imbalance 
in the case of finite $T$. 
Indeed, in Appendix C we show that fixing the condensate particle number  implies the transition is (within numerical error) unaffected by the presence of thermal particles for all barrier heights  probed. 


Furthermore, we note that the condensate dynamical regime across the Josephson junction depends on   the value $V_0/\mu(T=0)$, where $\mu(T=0)$ is the zero temperature chemical potential obtained by the Gross-Pitaevskii equation in the limit of all particles being in the superfluid. Given that $\mu(T=0)$ is held fixed throughout our simulations, all our analysis is thus done at fixed $V_0=104 \hbar \omega _x =0.97 \mu(T=0)$.

For completeness, we note here that the third dynamical regime, namely self-trapping, is not relevant to this work, as -- for our parameter set -- it manifests itself at a much  higher value of $V_0/ \mu(T=0) \simeq 1.7$.

In the subsequent analysis of Sec.~III we will consider those 2 regimes separately (Secs.~C and D  respectively) to focus on the role of the thermal cloud on particle dynamics.

\subsection{Second Order Josephson Junction Contributions}


Given that our study identifies more than one relevant superfluid oscillations frequencies both at $T=0$ and in the $T>0$ regimes, it is appropriate here to briefly summarize prior relevant work identifying such multiple frequencies and the role of their arising couplings \cite{Ic2,singh20,zaccanti_19,Xhani20,Kwon84,bidasyuk_16,Luick20,modugno2018,Goldobin2007}.

At first order in the tunneling Hamiltonian, only condensate-to-condensate tunneling term contributes to the superfluid current and the current-phase relation is sinusoidal $I=I_c \sin(\Delta \phi)$. This is valid as long as the barrier height is much larger than $\mu$. 
However, if instead $V_0 \sim \mu$, second order terms must be considered which originate from the tunneling between condensate and non-condensate states.
At $T=0$ and for a BEC the latter consists of phonon modes \cite{Ic2,MQST2}.

In the presence of a finite chemical potential difference between the two wells, this second order term could lead to the presence of an additional non-dissipative (of the form `$\sin (2\Delta \phi)$') and/or dissipative term (of the form `$\cos(2\Delta \phi)$') in the current-phase relation, with the latter being finite even at $T=0$. The presence and importance of the coherent (non dissipative) term oscillating at double the $\nu_J$ frequency has been studied in several papers both with bosonic and fermionic systems \cite{Ic2,zaccanti_19,singh20,Xhani20,Kwon84,Luick20}. Moreover, for our geometry, Ref.~\cite{Xhani20} shows that the profile of the maximum superfluid current flowing through the junction versus the barrier height $V_0/\mu$ could be described by the presence of both the first order (`$\sin(\Delta \phi)$') and the non-dissipative second order (`$\sin(2\Delta \phi)$') term in the current-phase relation, with the second  having a negative sign.
In fact, a recent study with a point-contact junction~\cite{Brantut2020} shows the presence even of a dissipative current term  in that geometry, as predicted by  \cite{Ic2}.
 
In the following sections we will show that our analysis, based on the long-time evolution of the superfluid dynamics, suggests that both dissipative and non-dissipative second order terms in the superfluid current (i.e.~population imbalance) could become important and their presence depends on the dynamical regime. For completeness, we also note here that the multimode regime is also found in the highly-excited self-trapping regime for an initial imbalance much larger than a critical value \cite{XhaniNJP,bidasyuk_16,modugno2018}.


\subsection{Self-Consistent Finite-Temperature Kinetic Model}

We model the system as the sum of a condensate and a thermal part, in the context of the collisionless Zaremba-Nikuni-Griffin (ZNG) formalism \cite{ZNG2,ZNG3,ZNG6,nick_book}.
 This technique, which has already been successfully applied to diverse non-equilibrium settings, including condensate growth~\cite{ZNG-growth}, collective modes~\cite{ZNG4,ZNG-scissors}, soliton~\cite{ZNG-soliton} and vortex \cite{vortexT1,vortexT2,vortexT3} dynamics is
described in more detail in Appendix~\ref{app_1}. 
The condensate  wavefunction $\psi$  evolves according to the generalized Gross-Pitaevskii equation
\begin{equation}
\label{gped}
i \hbar \frac{\partial \psi}{\partial t}=\left[ - \frac{\hbar^2 \nabla ^2}{2M}+V_\mathrm{ext}+g(|\psi|^2+2n_\mathrm{th})\right] \psi \;,
\end{equation}
which  accounts for the thermal cloud mean field potential, $2g n_\mathrm{th}$ \cite{ZNG2}. 
Here $M$ is the particle mass (here $^6$Li molecule), $V_\mathrm{ext}$ is the double-well potential defined above, $g=4 \pi \hbar ^2 a/M$ is the interaction strength with $a$ the corresponding $s$-wave scattering length, and $n_\mathrm{th}$ is the thermal cloud density. The condensate density is obtained from $n_{\rm BEC}=|\psi|^2$.
The thermal cloud dynamics are described through the phase-space distribution $f$ (where 
$n_{th}=1/(2\pi\hbar)^3 \int d\bold{p} ~ f(\bold{p},\bold{r},t)$), which satisfies the collisionless Boltzmann equation
\begin{equation}
\label{bolt}
\frac{\partial f}{\partial t}+ \frac{\bold{p}}{M} \cdot \nabla_{\bold{r}} f-\nabla_{\bold{r}}V_{\rm eff} ^{\rm th} \cdot \nabla_{\bold{p}}f=0
\end{equation}
where $V_{\rm eff}^{\rm th}=V_\mathrm{ext}+2g (n_{\rm BEC}+n_{th})$ is the generalized mean-field potential felt by the thermal particles, whose profile is shown in Appendix~\ref{app_2}.

Due to the repulsive interaction between condensate and thermal particles, the
thermal density $n_{th}$ is maximum where the
condensate density $n_{\rm BEC}$  is minimum. 
 This is evident in Fig.~\ref{cond_therm_dens_zcr} (b), where the thermal cloud (ii) is concentrated at the edges of the condensate density (i) and close to the barrier 
 where $n_{\rm BEC}$  is minimum.

Our numerical study for the superfluid, based on the techniques discussed in Refs.~\cite{ZNG2,ZNG3}, is conducted
in a grid of $\left[-24,24\right]l_x,\left[-4,4\right]l_x, \left[-4,4\right]l_x$ along the $x$, $y$ and $z$ directions respectively, where $l_x = \sqrt{\hbar / M \omega_x}$, based on $1024 \times 64 \times 64$ grid points for the condensate.
For the more spatially extended thermal cloud, we use
a corresponding double grid of size $\left[-48,48\right]l_x$, with 2048 grid points along the $x$ axis, further extended to $\left[-100,100\right]l_x$ and 2348 grid points for the highest probed temperatures $T \sim 0.8T_c$.
%
The broad temperature range  studied here corresponds to a condensate fraction
 $N_\mathrm{BEC}/N_\mathrm{tot}  \sim [0.1:1]$ (see Fig.~\ref{cond_therm_dens_zcr}(c)).

In this work we focus on the dynamical evolution of the condensate and total particle number fractional population imbalance across the Josephson and dissipative regimes, and analyze their dominant contributions and corresponding frequencies of oscillations, which display a range of interesting features.

\section{Dynamical regimes at Fixed Condensate Number  \label{sec3}}

\subsection{Key Parameters and Physical Variables}

In our main study we keep fixed the Gaussian barrier height $V_0=104 \hbar \omega _x$, and the condensate particle number  $N_{\rm BEC}=(5.04\pm0.02)\times10^4$, with the small error bar given by the maximum difference between the condensate number at different $T$ having no noticeable effect  on the characterised Josephson physics.
As a result, $V_0/\mu (T=0) =0.97$. This constraints the system dynamics to be either in the Josephson plasma oscillation (for small initial population imbalances), or in the dissipative regime (for larger population imbalances), thus staying far from the self-trapping regime. Moreover, smaller values of $V_0/\mu$ would imply being more in a hydrodynamical than a tunneling regime. Although earlier work has characterised in detail the generated vortex ring dynamics \cite{Xhani20,XhaniNJP}, our parameter choice here corresponds to a regime in which the generated vortex ring shrinks rapidly at the barrier location, making it  hard to directly visualize the vortex ring. Nonetheless, to confirm the existence of vortex rings at lower values of $V_0/\mu$, for which we have previously found them to be long-lived \cite{XhaniNJP}, we have done some analysis at $V_0/\mu \sim 0.6$ which clearly shows the generated vortex ring in the condensate as a region of locally reduced condensate density being infilled by the thermal cloud, consistent with \cite{vortexT1,vortexT2,vortexT3} -- see Appendix D for more details.
In order to also fix the initial condensate population imbalance when varying temperature, we use a fixed value of $\epsilon$ in the linear barrier shift contribution $-\epsilon x$ for each regime (Josephson, dissipative) studied.

We define the condensate $z_\mathrm{BEC}(t)$, the thermal cloud $z_\mathrm{th} (t)$ and the total $z_\mathrm{tot}(t)$ fractional population imbalances respectively as:
\begin{eqnarray}
z_\mathrm{BEC}(t)&=&\frac{N_R ^{BEC} (t)-N_L ^{BEC}(t)}{N_R ^{BEC} (t)+N_L ^{BEC}(t)}  \\
\nonumber \\
z_\mathrm{th}(t)&=&\frac{N_R ^{th}(t)-N_L ^{th}(t)}{N_R ^{th}(t)+N_L ^{th}(t)}  \\
\nonumber \\
z_\mathrm{tot}(t)&=&\frac{N_R(t)-N_L(t)}{N_R+N_L}=\frac{N_R(t)-N_L(t)}{N_{tot}}
\end{eqnarray}
where $N_{R/L} ^{BEC}$ and $N_{R/L} ^{th}$ are the number of the  condensate and thermal particles on the right/left sides of the barrier (centered at x=0) while $N_{R/L}$ is the sum of the number of the thermal particles and the condensate particles on the right/left sides of the junction; we also note that, by construction, the total particle number is conserved in our collisionless model.

Having fixed the ratio of $V_0/\mu$, the $T=0$ limit has a definitive value for the critical population imbalance marking the transition from plasma to dissipative regime; for our current parameters (and $N_{\rm BEC} \sim 50,400$), this occurs at $z_{\rm cr} ^{\rm BEC} =0.08$.
The system dynamics in the pure superfluid limit is thus determined by the sign (and magnitude) of
($z_0 ^{\rm BEC} -z_{\rm cr} ^{\rm BEC}$) \cite{Xhani20,XhaniNJP}.
In order to clearly analyse the role of temperature on the system dynamics, we choose to avoid potential transient issues very close to the dynamical transition point, and thus conduct our
analysis for two fixed values of $z_0$ chosen as $z_0 ^{\rm BEC}=0.046 < z_{\rm cr} ^{\rm BEC}$ (Josephson plasma regime) and $z_0=0.106 >z_{\rm cr} ^{\rm BEC}$ (vortex-induced dissipative regime), for which only a few vortex rings are generated. 

The presence of a thermal cloud introduces an additional relevant physical parameter for the system dynamics.
Specifically, the ratio $V_0/k_B T$ distinguishes between two dynamical regimes for the thermal cloud. For relatively low temperatures $V_0/k_B T \gtrsim 1$, the thermal cloud particles -- which would normally be constrained to either side of the barrier having insufficient energy to travel above it (exhibiting incoherent tunneling) -- can only propagate through their interaction with the condensed particles. However, in the opposite high-temperature regime $V_0/k_B T \lesssim 1$, the thermal particles have sufficient energy to overcome the barrier, and are thus allowed to execute oscillations in the underlying trap, hindered, but not precluded, by the Gaussian barrier forming the Josephson junction for the superfluid.    
As such, one would expect -- and we indeed find -- different dynamical behaviour to be dominating the low and high temperature regimes, observing a gradual change in the system dynamical behaviour around the regime  $T \sim V_0/k_B$.
In this work we probe the temperature range $k_B T/V_0 \in [0,3.1]$, corresponding to $T \in [0,220]$nK.
As our primary study keeps condensate particle number fixed with increasing temperature, this implies that the total particle number also increases with temperature, with the characteristic non-interacting critical temperature, $T_c$, thus being temperature-dependent, i.e.~$T_c(N(T))$.\\

We note that the chosen value $V_0$ of the barrier height used in our analysis (fixed by $V_0/\mu \sim 0.97$) corresponds for the simulated condensate particle number to an effective temperature $V_0/k_B \sim 70$nK.
Noting the changing {\em total} particle number and $T_c$ with temperature, our analysis is thus conducted in the range $T/T_c \in [0, 0.8]$, with the characteristic thermal energy separating the two thermal cloud dynamical regimes emerging (for the particular geometry and condensate particle number) at $V_0/k_B \sim 0.5 T_c$.

Having introduced our parameter choice, we now proceed to analyse the role of the thermal cloud on the dynamics in each regime, paying particular attention to the dependence of the plasma frequency on temperature, and the relative importance of this frequency on the system dynamics. 

\subsection{$T>0$ Josephson plasma Regime}
In the deep tunneling regime, the oscillation frequency $\omega _J$ of $z(t)$ depends on the characteristic Josephson junction energies, such as the tunneling energy $E_J$ and the onsite interaction energy $E_c$, with $\omega _J =(1/\hbar) \sqrt{E_c E_J}$ in the two-mode model approximation \cite{MQST1,andrea2003,MQST2,MQST3}. However for our barrier height and width ($V_0 \lesssim \mu$ and $w \simeq 4 \xi$) it has been shown \cite{Xhani20} that the two-mode model does not predict well the Josephson frequency.
Thus, our subsequent plots extract the relevant frequencies
from sinusoidal fits of the numerically-evaluated time evolution of the condensate imbalance (see following section). 

We start by analyzing the dependence of the system dynamics on temperature in the Josephson regime, upon fixing the initial {\em condensate} population imbalance $z_0^{\rm BEC}=0.046 < z_{\rm cr}^{\rm BEC}$ (with $z_0^{\rm BEC} / z_{\rm cr}^{\rm BEC} \sim 0.6$).

\begin{figure}[]
\centering
  \includegraphics[width=0.8\columnwidth]{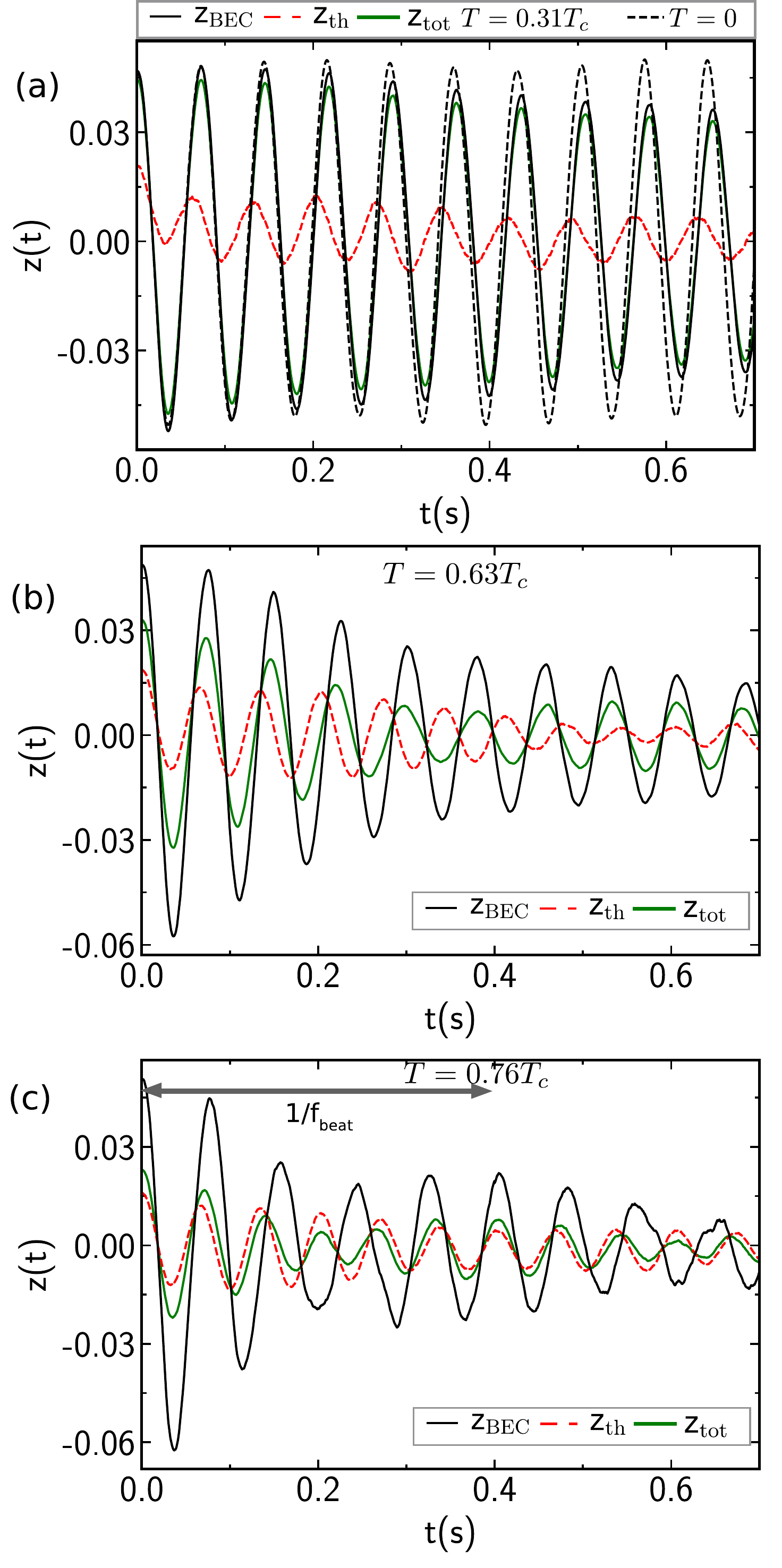}
  \caption{
  Population imbalance oscillations exhibiting damping (and, in some cases, beating) for the condensate (black), thermal (red) and total (green) population imbalances in the Josephson  regime ($z_0< z_\mathrm{cr}= 0.08$) at different temperatures:
(a)    $T=$ 40 nK $= 0.31 T_c$, 
(b) $T=$ 100 nK $= 0.63 T_c$,
and (c) $T=$ 160 nK $= 0.76 T_c$, where $T_c$ corresponds to the non-interacting critical temperature 
for the case of a fixed condensate number $N_{BEC} = 5.04 \times 10^4$ (such that $T_c$ varies with $T$). These data corresponds to a Gaussian barrier with $V_0=104 \hbar \omega _x = 0.97 \mu(T=0)$ and $w = 3.8\xi$.
For comparison, dashed black line in (a) depict the corresponding undamped single-frequency Josephson plasma oscillations in the pure superfluid ($T=0$) limit, which corresponds to the black dashed line in Fig.~1, i.e.~$z_0=0.046$. 
  }
  \label{fig:imb_jos}
\end{figure}

The evolution of the fractional relative population imbalance is shown in Fig.~\ref{fig:imb_jos} for (a) a low, (b) intermediate and (c) a relatively high temperature.
At low temperatures the small thermal fraction (red) is moved by the condensate motion, with no significant distinction between the condensate (solid black) and total (solid green) fractional population imbalances, both of which are slightly damped through the mutual friction of the condensate in its motion through the thermal cloud: such damping is evident in Fig.~\ref{fig:imb_jos}(a), which also shows -- for comparison -- the previously considered $T=0$ undamped plasma oscillations (dashed black line already shown in Fig.~1(d)).

However, as the temperature increases to values $k_B T > V_0$ [Fig.~\ref{fig:imb_jos} (b)-(c)], the increasing thermal component is free to execute its own oscillations over the barrier, at a distinct frequency to that of condensate oscillations.
As a result, the condensate oscillations are significantly damped, and the total population difference oscillation features becomes more similar to those of the thermal imbalance, with the combination of the two distinct frequencies leading to the emergence of beating. The higher the temperature, the shorter the beating time is, i.e.~its corresponding frequency $f_{\rm beat}$ is larger, thus favouring its observation even within experimental times. The reason for that will be clearer in the following sections.

At low $T$ the thermal cloud mean kinetic energy $k_B T$ is not high enough for the thermal cloud to flow hydrodynamically through the barrier. Thus thermal particles can only perform  incoherent tunneling through the barrier \cite{MQST2,MQST4}.  For relatively low $T$ such that the thermal cloud mean kinetic energy $k_B T$ is smaller or comparable to $(V_0-\mu_0)$, the crossing rate of a thermal particle across the barrier is given by the Arrhenius-Kramer formula \cite{MQST2}:
\begin{equation}
P_{th} \simeq \frac{\omega _x}{2\pi} \rm{exp}\left( -\frac{(V_0-\mu_0)}{k_B T} \right)
\label{prob_th}
\end{equation} 
 where $\mu _0=\mu (N/2)$. In the case of $T=0.31 T_c$, shown in Fig.~\ref{fig:imb_jos}(a), Eq.~(\ref{prob_th}) estimates a crossing rate $P_{th}\sim 10$Hz, which corresponds to a time $\tau_{th} \sim 0.1$s. This means that each $\sim 0.1$s, thermal particles may cross the barrier via tunneling. Such result is consistent with the $z_{th}(t)$ profile in  Fig.~\ref{fig:imb_jos}(a)(red profile) which oscillates initially around a non-zero mean value $\langle z_{th}(t)\rangle$ for at least $\sim 0.1$s with its mean value decaying in time until it achieves a value near zero.

Closer inspection of the superfluid oscillations reveals contributions from more than one frequency, even at low temperatures.
This might have been expected as in our parameter regime ($V_0/\mu \simeq 1$ and  $w/\xi \simeq 4$), which is not in the deep tunneling regime, the superfluid current (thus the condensate imbalance) is expected to oscillate with two frequencies; a dominant Josephson plasma frequency  and an additional second frequency arising from the second order-term in the tunneling amplitude \cite{Ic2}.

Moreover, at finite $T$, the presence of the thermal cloud  introduces a normal component $I_n$ to the Josephson current (an Ohmic-like contribution) \cite{MQST2,Ic2} $I_n =-G_{th} \Delta \mu$ where $G_{th}$ is the junction conductance. In the limit of low $T$, the latter can be estimated from the Arrhenius-Kramers formula as $G_{th}=P_{th} N_{th}/k_B T$  \cite{MQST2}, with $\Delta \mu$ the chemical potential difference between the two wells being proportional to the condensate imbalance (i.e. $\Delta \mu(t) \propto z_{\rm BEC}(t)$). For the chosen barrier height, the Arrhenius-Kramers formula  is valid for very low temperature  as the barrier height here is not very different from the single-well chemical potential. 
Based on these two considerations and knowing that the condensate current is related to the condensate imbalance through $I=-(N_{\rm BEC}/2)\ dz_{\rm BEC}/dt$, we then choose as our fitting function for the condensate imbalance a function of the form: 
\begin{equation}
\begin{split}
F(t)& = \, a_J \, \cos(2\pi\nu _J t+\phi _J)\, \exp(-\gamma _J t)  \\
& + \, a_i \, \cos(2\pi\nu_i  t+\phi_i) \, \exp(-\gamma_i  t)
\label{fits-j}
\end{split}
\end{equation}
which, in addition to distinct amplitudes ($a_{.}$), frequencies ($\nu_{.}$) and damping rates ($\gamma_{.}$) for the two components [where the dot $.$ subscript denotes either $J$, or $i$], also allows for unconstrained phases $\phi_{.}$ of each contribution. 

 \begin{figure*}[t!]
\centering
  \includegraphics[width=\textwidth]{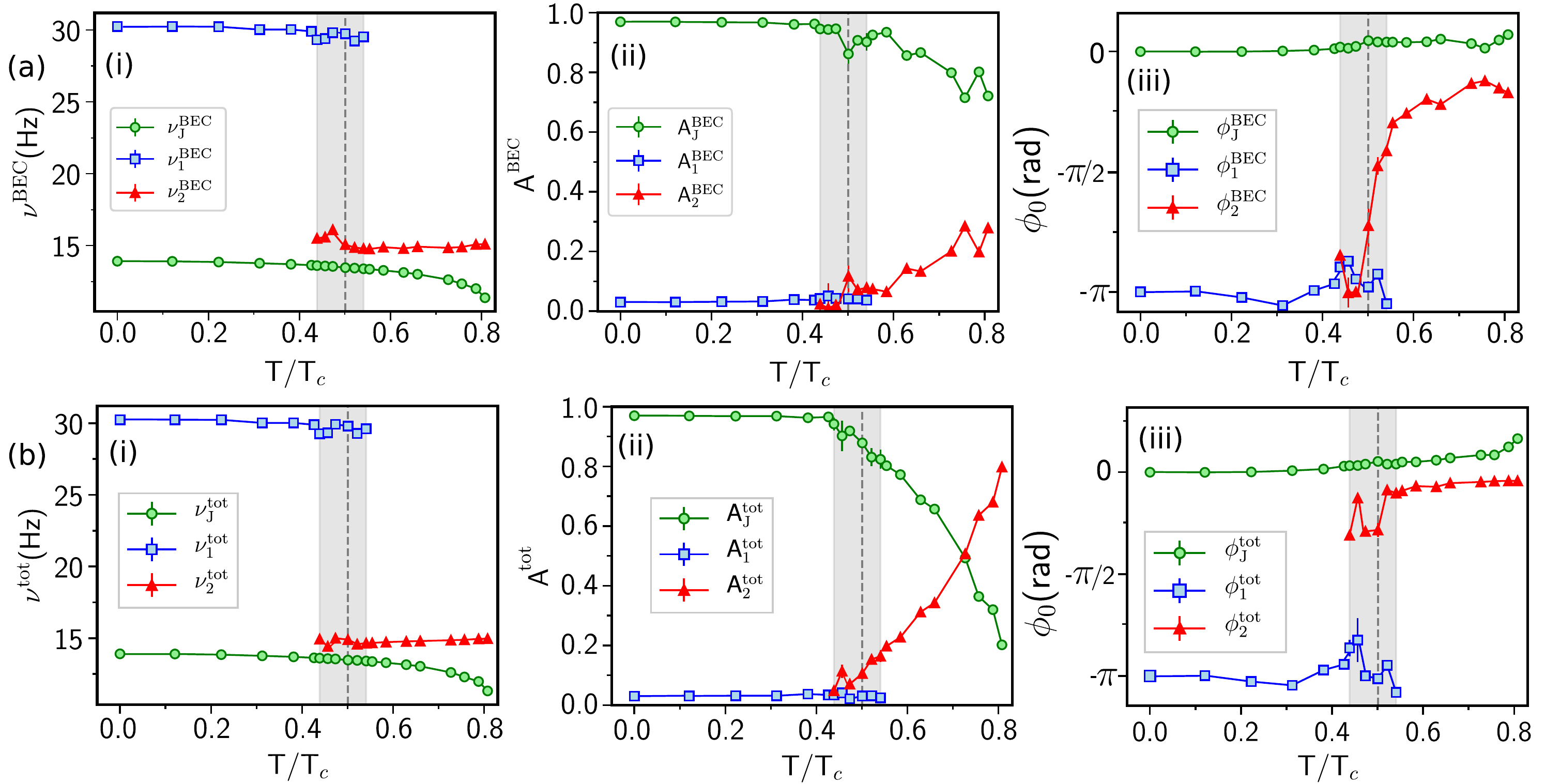}
  \caption{Temperature dependence of oscillatory dynamics of (a) the condensate, and (b) the total population imbalance across a Josephson junction for a system of fixed condensate particle number, barrier amplitude $V_0/\mu \sim 0.97$, and  initial condensate imbalance $z_0<z_\mathrm{cr}$ (such that the pure superfluid dynamics are in the Josephson dynamical regime). 
  Shown in each case are the (i) dominant oscillation frequencies, (ii) their relative amplitudes (extracted from  Eq.~(\ref{fits-j})-(\ref{amplitudes-j})), and (iii) their corresponding initial phases (extracted through the fits of Eq.~(\ref{fits-j})) as a function of scaled temperature.
  The vertical dashed lines indicate the temperature corresponding to the barrier height, i.e.~$T=V_0/k_B=70 nK$ which corresponds, for the chosen parameters, to $T/T_c \sim 0.5$, with the grey band indicating the region around this characteristic temperature in which there is a transition in the relative importance of different frequency contributions. 
  Population imbalances are fitted in the time interval $[0.05:0.72]$s either by two-frequency fits when two components are clearly predominant (outside the grey band), or by a combination of two- and three-frequency fits in the intermediate region (within the grey band): in the latter case, depicted values and error bars are extracted by averaging over values obtained by the independent two- and three-frequency fits. 
  }
  \label{fig:nu_ampl_all_jos_bec}
\end{figure*}

Firstly we focus on the condensate  $z_\mathrm{BEC}(t)$ oscillations, and their dependence on temperature. Examination of both Fourier Transforms and the above functional fits reveal, as expected, periodic oscillations at a dominant frequency -- which we interpret as the Josephson plasma frequency, labelled by the subscript $J,$ 
along with a clearly-identified secondary frequency over the entire temperature range probed.

Interestingly, the secondary oscillations (labelled here by $i=1,2$) correspond to distinct frequencies at low  temperatures $k_B T/V_0 \lesssim 1$ (henceforth labeled as $\nu_1$) and high  temperatures $k_B T / V_0 \gtrsim 1$ (henceforth labelled as $\nu_2$).
In order to capture the transition from $\nu_1$ to $\nu_2$ with increasing temperature, and avoid introducing any bias to our results, the analysis in a narrow region around $k_B T \sim V_0$ is  extended to 3-frequency fits ($\nu_J$, $\nu_1$ and $\nu_2$) to provide some continuity to our analysis. 


Beyond characterizing the oscillation frequencies and damping rates, we also investigate 
the relative contributions, $A_{.}$, of each component defined for the 2-component fits as:
 \begin{eqnarray}
A_{J}=\frac{a_J}{a_J+a_i} \hspace{1.0cm} {\mathrm and} \hspace{1.0cm}
A_i=\frac{a_i}{a_J+a_i}
\label{amplitudes-j}
\end{eqnarray}
(with $a_i$ in the denominator replaced by $\sum_{i=1,2} a_i$ for the case of 3-frequency fits).
Finally we investigate the phases $\phi_{.}$ of the different contributions.

Such information is plotted in Fig.~\ref{fig:nu_ampl_all_jos_bec} for both (a) the condensate fractional population imbalance $z_{BEC}(t)$ [top plots], and (b) the corresponding total fractional imbalance $z_{\rm tot}(t)$ [bottom panels]. In each case we show the dominant frequencies (left column, (i)), their relative contributions (middle, (ii)) and their individual phases (right, (iii)).

Let us now analyze our findings, focussing initially on the $z_{BEC}(t)$ oscillation frequencies [Fig~\ref{fig:nu_ampl_all_jos_bec}(a)]: At low $k_B T \ll V_0$ the dominant Josephson dynamics (labelled by green circles) occurs at the frequency $\nu_J \approx 14$Hz [panel (i)], with a relative weighting exceeding 97\% [panel (ii)], and occurring -- as expected -- without any initial phase delay, i.e.~$\phi_J^{\rm BEC}=0$ [panel (iii)]. As temperature increases towards $V_0$ (corresponding here to a condensate fraction reduction of $\sim 30\%$), the Josephson frequency exhibits a small monotonic decrease on the few \% level.
The low-temperature frequency $\nu_1 \approx 30$Hz (blue squares) detectable thus far with a relative amplitude of few $\%$ and a phase offset of $\sim \pi$, is gradually supplemented by an  additional frequency $\nu_2 \approx 15$Hz which becomes dominant as temperatures increase beyond  $k_B T / V_0 \sim 1$ (indicated by the vertical dashed grey line). At the highest temperature probed here ($k_B T / V_0 \sim 3.1$ , $T/T_c = 0.81$), with $N_{\rm BEC}/N_{\rm tot} = 0.1$, the Josephson plasma frequency contribution to the condensate fractional imbalance decreases to about 70\%, with the relative phase difference $| \phi _{\rm J} ^{\rm BEC} - \phi _{2} ^{\rm BEC}| \lesssim \pi/4$.
We note that the presence of the $\nu _1 \approx 2 \nu_J$ frequency component in $z_{\rm BEC}(t)$ even  at $T=0$ with an initial phase $-\pi$ with respect to the $\nu _J$ component is consistent with the presence of a second-order (non-dissipative) term with an opposing (negative) sign in the current-phase relation. The contribution of such a double Josephson plasma frequency oscillation term~\cite{Ic2,Xhani20} was previously
%
found to be  only few percent for $V_0 \simeq \mu$ \cite{zaccanti_19},
consistent with our current picture.

\begin{figure*}[t!]
\centering
  \includegraphics[width=.99\textwidth]{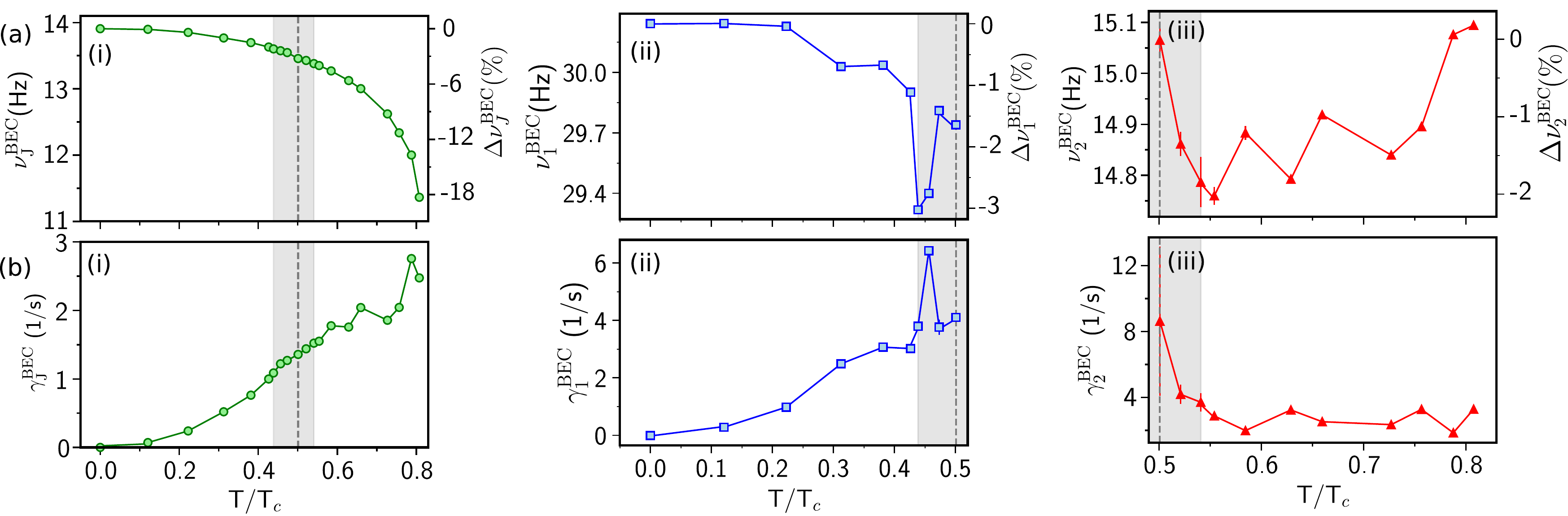}
  \caption{
  Temperature dependence of (a) dominant frequencies, and (b) damping rates of the {\em condensate} component corresponding to the parameters of Fig.~3(a).
  These are shown for (i) the dominant, Josephson plasma, frequency $\nu_J^{\rm BEC}$ [plotted over entire temperature range], and (ii)-(iii) the other arising secondary frequencies over their respective regions of importance, i.e.~(ii) $\nu _1^{\rm BEC}$, corresponding to the frequency around 30 Hz, and (iii) $\nu _2^{\rm BEC}$ corresponding to the frequency around 15 Hz (which is also the trap frequency along the $x$ axis).
  The right axes of the frequency plots in (a)  also show the $\%$ change of each frequency, which reveal a notable $18\%$ lowering for $\nu_J^{\rm BEC}$, but a much weaker, few $\%$, dependence for $\nu_1^{\rm BEC}$ and $\nu_2^{\rm BEC}$.
  (b) Josephson frequency damping rates [(i), $\gamma_J^{\rm BEC}$] increase monotonically with increasing temperature, and remain moderate over entire probed regime, compared to $\gamma_1^{\rm BEC}$ and $\gamma_2^{\rm BEC}$ which respectively increase with increasing/decreasing temperature as they approach the crossover temperature $T \sim V_0/k_B$, labelled by the vertical dashed line.
  %
}
  \label{fig:nu_main_bec_jos}
\end{figure*}

The above analysis was based entirely on the {\em condensate} motion, and the back action that the thermal cloud has on it. Although we have also separately analyzed the thermal cloud population imbalance dynamics, a more complete picture of the coupled system dynamics can be obtained by looking at the imbalance of the total population, $z^{\rm tot}(t)$, with results  shown in Fig.~\ref{fig:nu_ampl_all_jos_bec}(b).
Again, the same 3 frequencies are found ($\nu_J$, $\nu_1$ and $\nu_2$) appearing in the same temperature ranges, but there is a critical difference: in our simulations,  based on keeping the condensate particle number fixed, the increase in temperature leads to an increase in the number of the thermal cloud particles: as such, while the Josephson frequency is not significantly affected, its relative contribution decreases rapidly as the total population becomes more dependent on the increasing thermal contribution:  at $T/T_c \sim 0.81$ the Josephson plasma mode contributes only about 20\% of the total amplitude, while the phases of the two dominant contributions approach each other, indicating initial phase-locking. The increasing relative contribution of $\nu _2$ with $T$, combined with the fact that the value of $\nu_2$ is close to that of $\nu_J$, causes the relative total population imbalance to exhibit beating between these two components with a beating frequency given by $f_{beat}=\vert \nu _2 ^{tot}-\nu _J ^{tot}\vert$: its inverse identifies a characteristic beating timescale, as indicated by an arrow in the $z_{tot}(t)$ profile in Fig.~\ref{fig:imb_jos}(b). The beating frequency increases with higher $T$, as the $\nu _2 ^{\rm tot}$ value tends towards the trap frequency while the $\nu _J ^{\rm tot}$ becomes even smaller. Moreover, we note that $f_{\rm beat}$ becomes larger even in the case of fixed $T$ and larger barrier height as $\nu _J$ decreases with $V_0$ \cite{MQST2}. Thus the beating effect, which is a consequence of the effect of the thermal cloud, could be visible at shorter time ($\tau_{\rm beat}=1/f_{\rm beat}$) for relatively large $T$ in case of fixed $V_0$ or for relatively large barrier height (but still not in the self-trapping regime) at fixed $T$.

More information on the properties of the dominant contributions to $z_{\rm BEC}(t)$ can be found in Fig.~\ref{fig:nu_main_bec_jos} which focuses on the dependence of the frequencies $\nu_{\rm J}$, $\nu_{1}$ and $\nu_{2}$ [top row] and corresponding damping rates [bottom row] as a function of scaled temperature $T/T_c$. We clearly see the monotonic decrease of $\nu_{\rm J}^{\rm BEC}$ over the entire temperature range probed, accompanied by a super-linear increase of the damping.
Frequency $\nu_1$ also decreases with increasing temperature, and its contribution damps at a faster rate than the $\nu _{\rm J}$-term, whereas $\nu_2$ displays a less clear dependence and a corresponding large damping which increases with decreasing temperature, eliminating that mode for $k_B T \ll V_0$: this somewhat counter-intuitive behaviour can be understood from the fact that at such lower temperatures, the thermal cloud cannot on its own move across the barrier, but can only do so mediated by the condensate which drags it along. Since the analysis here focuses on behaviour extracted from the condensate imbalance dynamics, and the small thermal component does not drag the condensate at low temperatures, it is understandable that no $\nu_2$ contribution can be found at such low temperatures. 

\begin{figure}[t!]
\centering
  \includegraphics[width=.6\columnwidth]{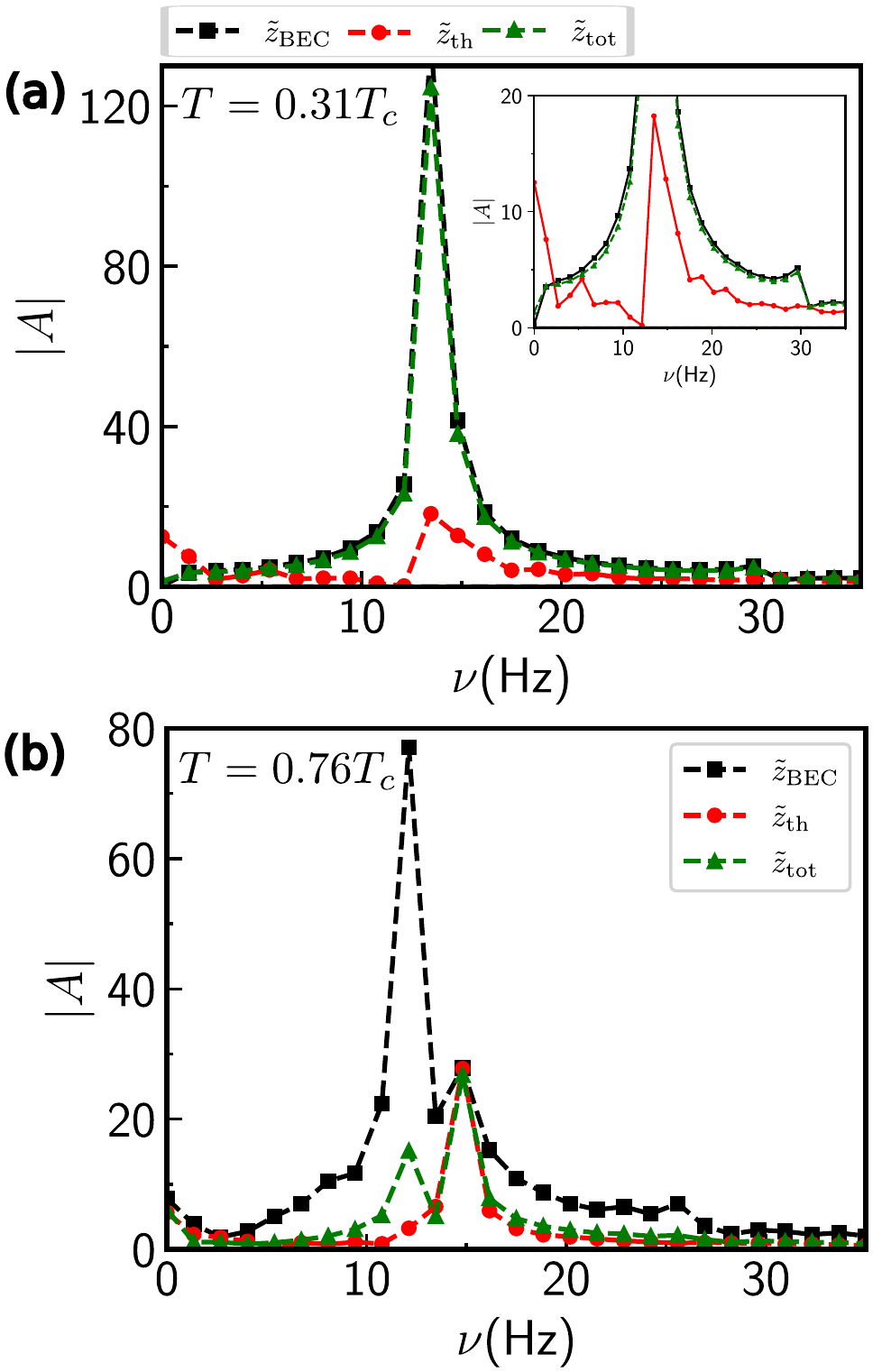}
  \caption{The Discrete Fourier transform (DFT) of the different population imbalance dynamics corresponding to the parameters of Fig.~3 ($z ^\mathrm{\rm BEC} _0 <z_\mathrm{cr} ^\mathrm{\rm BEC}$) at two temperatures chosen such that (a) 
  $T=$40nK$=0.31T_c \ll V_0 / k_B$ (a) and (b) $T=$160 nK$= 0.76 T_c \gg V_0/k_B$.
  Depicted in each case are the contributions arising from the oscillating dynamics of the condensate
  $z_\mathrm{\rm BEC}$(t) (black line), the thermal component $z_\mathrm{\rm th}$(t) (red) and the total population imbalances $z_\mathrm{\rm tot}$(t) (green).
  The $y$ axis is the amplitude of the components of the DFT $\vert \tilde{z}\vert$ while the $x$ axis is the frequency, exhibiting a numerical resolution of 1.3Hz.
  The inset in (a) plots a zoomed-in version which reveals the importance of the emerging $\nu_1$ frequency across the different components analysed.
  } 
  \label{fig:dft_th_jos}
\end{figure}
These results are also confirmed by examining the Discrete-Fourier Transform (DFT) of the condensate, thermal and total population imbalance time series defined as:
\begin{equation}
\tilde{z}(\omega _m)= \sum _{j=0} ^{N} z(t_j) e ^{-i \omega _m t_j}
\end{equation}
where $\omega _m =2 \pi m/t_m$ which are shown in Fig.~\ref{fig:dft_th_jos} for (a) a low temperature, $T=0.31T_c \ll V_0/k_B$, and (b) a much higher one, $T=0.76T_c \gg V_0/k_B$. 
In the former case [Fig.~\ref{fig:dft_th_jos}(a)], the DFT shows a large amplitude peak at $\nu _J$ and a very small component at $\nu _1$ which can be just resolved in the appropriate zoomed-in plot. 
As a consequence, even the thermal imbalance spectrum shows a component at the dominant condensate frequency  $\nu_J$. At high $T$ instead the thermal imbalance spectrum's main frequency is close to $\nu _2$, i.e.~around 15Hz, which is different from the Josephson frequency $\nu_J$. Moreover, the total imbalance (light green) has two main components, one close to the dominant  condensate frequency  and one close to the thermal imbalance main component and due to their comparable relative contributions, the $z_{\rm tot}$ shows beating between these two frequencies at such high $T$. 


\subsection{$T>0$ Vortex-induced Dissipative Regime}

We now consider the temperature dependence of the dynamics in the other dynamical regime of the junction, namely the vortex-induced dissipative regime \cite{Xhani20,XhaniNJP}, induced by an initial condensate population imbalance $z_0^{\rm BEC}$ which exceeds the corresponding critical value for plasma oscillations, i.e.~$z_0^{\rm BEC} > z_{\rm cr} ^{\rm BEC}$.
As previously discussed, in the pure $T=0$ superfluid limit, under such conditions the ensuing dynamics is associated with the emission of one (or multiple consecutive) vortex rings   and associated acoustic emission \cite{Xhani20,XhaniNJP}, which causes a rapid decay of the condensate imbalance 
during its first quarter-cycle of uni-directional flow .

\begin{figure}[h!]
\centering
  \includegraphics[width=0.7 \columnwidth]{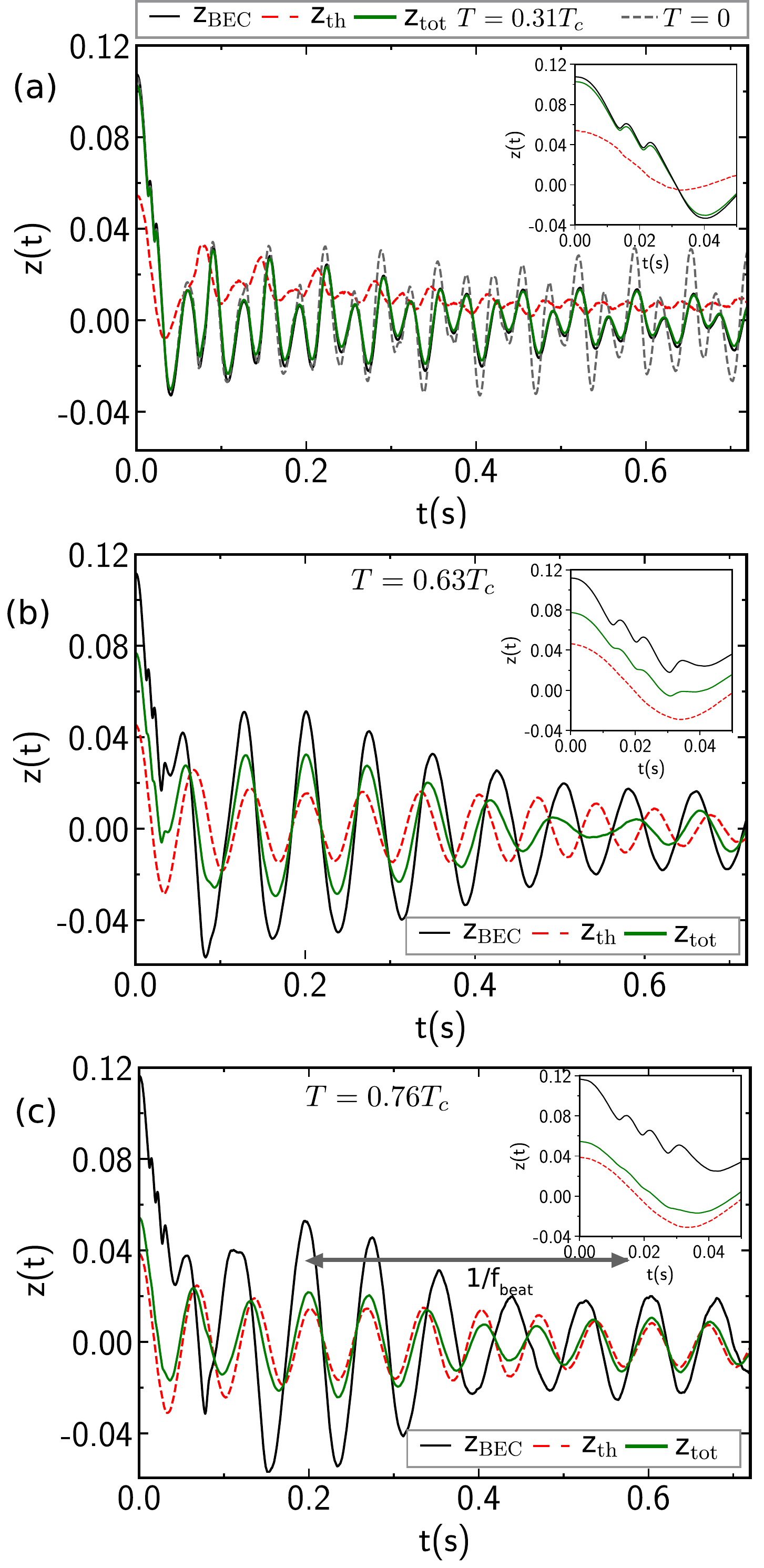}
  \caption{
  Population imbalance oscillations for the condensate (black), thermal (red) and total (green) populations imbalance in the vortex-induced dissipative  regime (i.e.~for $z_0=0.11> z_\mathrm{cr} = 0.08$) at different temperatures:
(a)    $T=$ 40 nK $= 0.31 T_c$, 
(b) $T=$ 100 nK $=0.63 T_c$,
and (c) $T=$ 160 nK $= 0.76 T_c$
for the case of a fixed condensate number $N_{BEC} = 5.04 \times 10^4$ and a Gaussian barrier with $V_0=104 \hbar \omega _x = 0.97 \mu(T=0)$ and $w = 3.8\xi$.
Corresponding insets zoom into the early time behaviour to reveal the characteristic kinks in the {\em  condensate} dynamics consistent with vortex generation; the importance of such dynamics in the total population imbalance clearly decreases with increasing temperature.
The plots in (a) also show the pure superfluid ($T=0$) results by dashed grey line, revealing that the vortex ring generation process occurring at {\em early} times (inset to (a)) at low temperatures is practically indistinguishable from the corresponding $T=0$ results, although the coupling to the thermal cloud induces more damping at {\em later} times.
Beating emerges already for the case considered in (b), but becomes clearly pronounced [as indicated] on the probed timescale in (c).
%
}
  \label{fig:imb_diss}
\end{figure}

The effect of temperature on such dynamics can be seen through
characteristic (a) low, (b) intermediate and (c) high temperature plots of the corresponding condensate/thermal/total fractional population imbalances shown in Fig.~\ref{fig:imb_diss}.

\begin{figure*}[t!]
\centering
  \includegraphics[width=.99\textwidth]{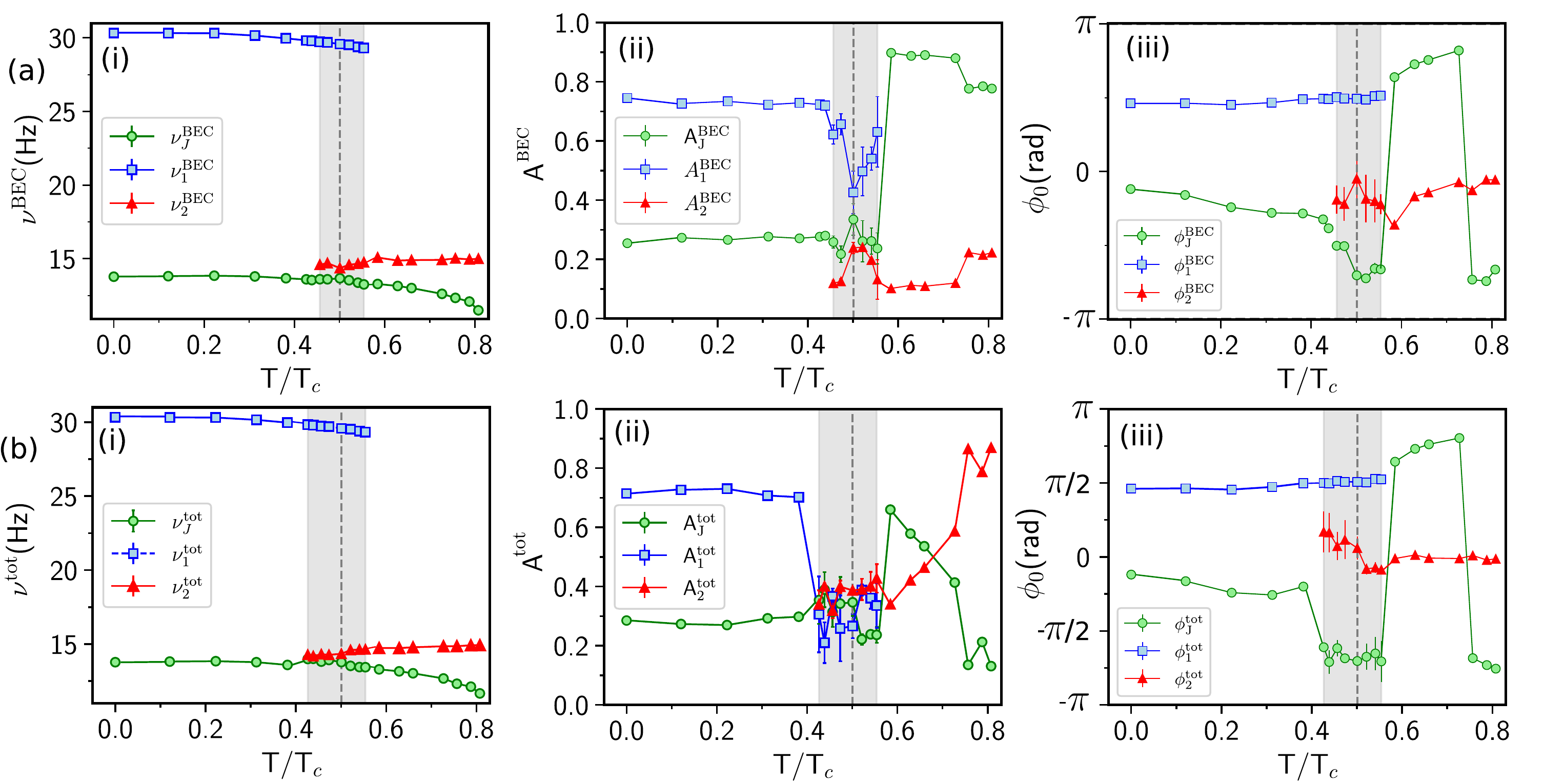}
  \caption{
  Temperature dependence of oscillatory dynamics of (a) the condensate, and (b) the total population imbalance across a Josephson junction for a system of fixed condensate particle number, barrier amplitude $V_0/\mu \sim 0.97$, and  initial condensate imbalance $z_0 > z_\mathrm{cr}$. 
    Shown in each case are the (i) dominant oscillation frequencies, (ii) their relative amplitudes (extracted  Eq.~(\ref{fits-j})-(\ref{amplitudes-j})), and (iii) their corresponding initial phases (extracted through the fits of Eq.~(\ref{fits-j})) as a function of scaled temperature. 
   Figure is closely related to the earlier Fig.~3, with all parameters/analysis/plotted lines having the same meaning,  except that here $z_0 > z_\mathrm{cr}$, so that the superfluid is in the vortex-induced dissipative regime (as opposed to the Josephson dynamical regime considered in Fig.~3).
}
  \label{fig:nu_diss_bec_tot}
\end{figure*}

At low temperatures where there is only a small thermal component, the total population imbalance is again dominated by the corresponding condensate one [Fig.~6(a)];
as such, they both reveal
the characteristic early-time signature of vortex ring generation (black/green curves), through the two kinks in the early dynamical evolution of $z_{\rm BEC}(t)$, around 15ms and 20ms (see also zoomed-in plot); although such vortex generation dynamics is practically indistinguishable from the pure superfluid $T=0$ case [shown by dashed grey line] differences do arise in the longer-term evolution, in the form of thermally-induced damping.
In this limit, the small thermal cloud dynamics is largely due to the condensate motion, 
with a small phase shift between them
due to the repulsive interaction between condensate and thermal particles. 
At such low $T$, the thermal particles cannot pass over the barrier, so they exhibit incoherent tunneling. In the vortex-induced dissipative regime the thermal imbalance oscillates around a non-zero value for longer time with respect to the Josephson plasma regime. This could be understood by noticing that in the former vortex rings are generated at the barrier position whose core is filled by the thermal cloud, i.e. it acts like a local `trapping' potential for the thermal particles, making it even harder for thermal particles to tunnel through the barrier.  
We note that for the chosen barrier height value the vortex rings shrink within the barrier region without propagating \cite{XhaniNJP}.
After the shedding of the generated vortex rings, the condensate and therefore the total imbalance oscillate about a zero mean value at two frequencies (which will be shown to correspond to $\nu_J$ and $\nu_1$), exhibiting  damping.

As the temperature increases to values higher than the barrier [already visible in Fig.~6(b)], the thermal cloud (red line) exhibits its own dominant oscillatory decaying dynamics across the barrier. Interestingly, the condensate mode exhibits enhanced damping (due to the relative motion through the dynamical thermal cloud) and very quickly the condensate starts oscillating with a single dominant frequency, driven by the oscillating thermal component. The total imbalance profiles becomes more similar to the thermal one as the thermal fraction (i.e.~as $T$) increases, and the presence of kinks during its initial decay becomes less visible in them with increasing $T$.

An interesting emerging feature here is the appearance of a third kink in the condensate population dynamics in the high-temperature region [Fig.~6(c) and corresponding inset], i.e.~a third vortex ring is generated. 
We can trace this back to a small shift in $z_{\rm BEC}(t=0)$ with temperature: even though our temperature-dependent analysis fixed linear tilted potential parameter $\epsilon$ and condensate particle number, such value of $z_{\rm BEC}(t=0)$ is indirectly affected by the fact that the thermal component dominates at the edges of the condensate density, thus  slightly reducing (through mean field repulsion) the condensate extent (and thus volume).
%
%
We have indeed confirmed that if instead of fixing $\epsilon$ with varying temperatures, we had explicitly chosen to fix $z_{\rm BEC}(t=0)$ in this $T>0$ case to exactly the same value as for $T=0$, the number of vortex rings being generated initially would be practically identical, even though differences would then emerge in the subsequent condensate dynamics. 
This is further discussed in Appendix~\ref{app4}. Furthermore, we also note that the early-stage dynamics of the condensate imbalance (the initial decay) is slightly affected by the thermal particles, with the main dissipative mechanism being the generation/dynamics of vortex rings and  associated sound waves.


\begin{figure*}[t!]
\centering
  \includegraphics[width=.99\textwidth]{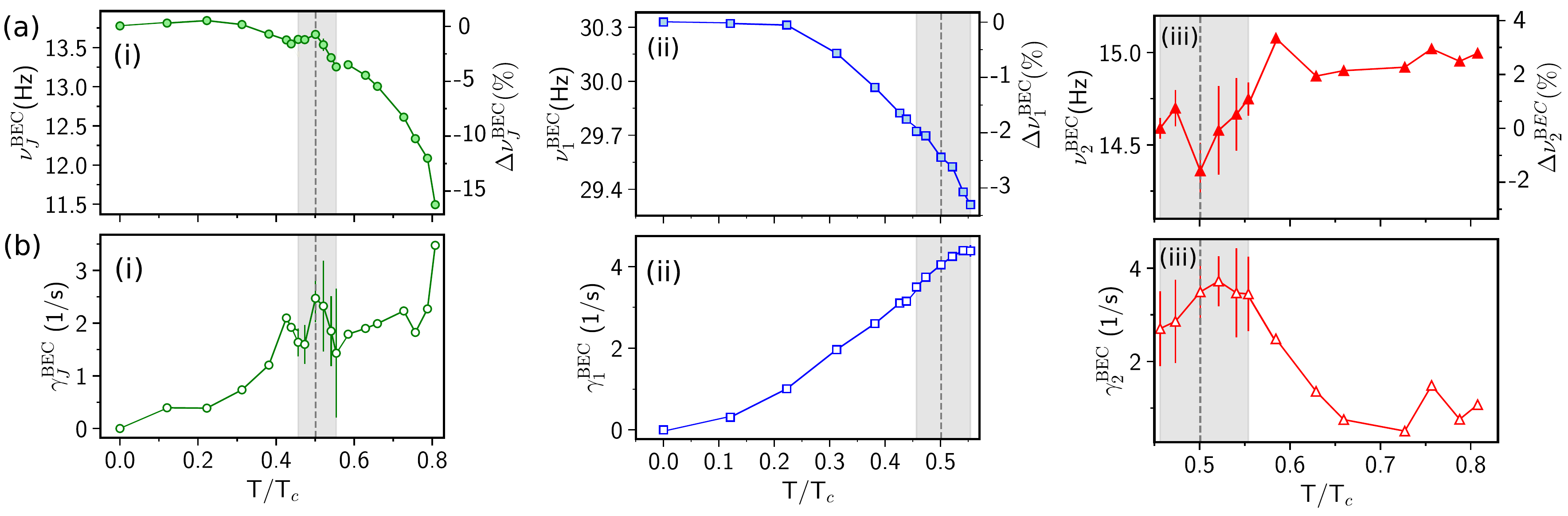}
  \caption{
   Temperature dependence of (a) dominant frequencies, and (b) damping rates of the {\em condensate} component corresponding to the parameters of Fig.~7(a)
  These are shown for (i) the Josephson plasma, frequency $\nu_J^{\rm BEC}$ [plotted over entire temperature range, but sub-dominant contribution for $T \lesssim V_0/k_B$], (ii)-(iii) the other arising frequencies over their respective regions of importance: 
  these are (ii) $\nu _1^{\rm BEC}$, corresponding to the frequency around 30 Hz which becomes dominant in the low-temperature regime due to the importance of acoustic emission during superflow dissipation and (iii) $\nu _2^{\rm BEC}$ corresponding to the frequency around 15 Hz (which is also the trap frequency along the $x$ axis).
  The right axes of the frequency plots in (a)  also show the $\%$ change of each frequency, which reveal a notable $16\%$ lowering for $\nu_J^{\rm BEC}$, but a much weaker, few $\%$, dependence for $\nu_1^{\rm BEC}$ and $\nu_2^{\rm BEC}$.
  (b) Josephson frequency damping rates [(i), $\gamma_J^{\rm BEC}$] increase monotonically with increasing temperature, while $\gamma_1^{\rm BEC}$ and $\gamma_2^{\rm BEC}$ respectively increase with increasing/decreasing temperature as they approach the crossover temperature $T \sim V_0/k_B$, labelled by the vertical dashed line.
}
  \label{fig:nu_diss_bec}
\end{figure*}


At relatively high $T$ [subplots (b)-(c)], the condensate and total imbalance show beating, whose frequency $f_{\rm beat}$ (period $\tau_{\rm beat}$) becomes larger (smaller) at higher $T$. 
%
%
An analysis (similar to Fig.~\ref{fig:nu_ampl_all_jos_bec}) of the dominant frequencies, relative contributions and initial phases as a function of temperature is shown in Fig.~\ref{fig:nu_diss_bec_tot} for both (a) the condensate fractional population imbalance oscillations $z_{\rm BEC}(t)$ (top) and (b) the total population imbalance $z_{\rm tot}(t)$ (bottom). In the vortex-induced dissipative regime the fit is performed after the initial decay.


Remarkably, the same 3 frequencies emerge, as found previously in the case of the Josephson regime both for the condensate (Fig.~\ref{fig:nu_diss_bec_tot}(a)(i)) and total imbalance  (Fig.~\ref{fig:nu_diss_bec_tot}(b)(i)).
However, an important distinction becomes immediately apparent: although the dominant frequencies are the same as before, their relative contributions and initial relative phases of oscillations are not.
Specifically, Fig.~\ref{fig:nu_diss_bec_tot}(a)(ii) and  Fig.~\ref{fig:nu_diss_bec_tot}(b)(ii) shows clearly that the Josephson plasma frequency term is no longer dominant at low temperatures, contributing less than 30\% to the total amplitude at low temperatures in both condensate (a) and total imbalance (b).
The generation of vortices and sound waves leads to significant interaction between the condensate and its excited sound waves which affects the condensate imbalance spectrum even at $T=0$.  We note here that the frequency $\nu _1$ is related to the presence of sound waves as it will be evident in the following section. Moreover, the initial phase of such component is around $\pi/2$ shifted with respect to the $\phi _J$ (at relatively low $T$). This would cause the presence of a dissipative component at the condensate current which is finite even at $T=0$, consistent with the results in the paper \cite{Ic2}.

Once again the behaviour changes around $k_B T / V_0 \sim 1$, where the frequency $\nu_2$ emerges, due to the oscillations of the increasing thermal cloud in the underlying axial harmonic trap.
Interestingly, the primary role of the thermal cloud on the condensate motion initially appears to be to damp out 
the acoustic component with frequency $\nu_1$: thus, perhaps somewhat counter-intuitively, at higher temperatures the condensate reverts to single-frequency plasma oscillations, and so the Josephson contribution becomes more important, and the dominant ($\sim 80\%$) contribution to the condensate imbalance oscillations at higher temperatures. 

\begin{figure}[h!]
\centering
  \includegraphics[width=.6\columnwidth]{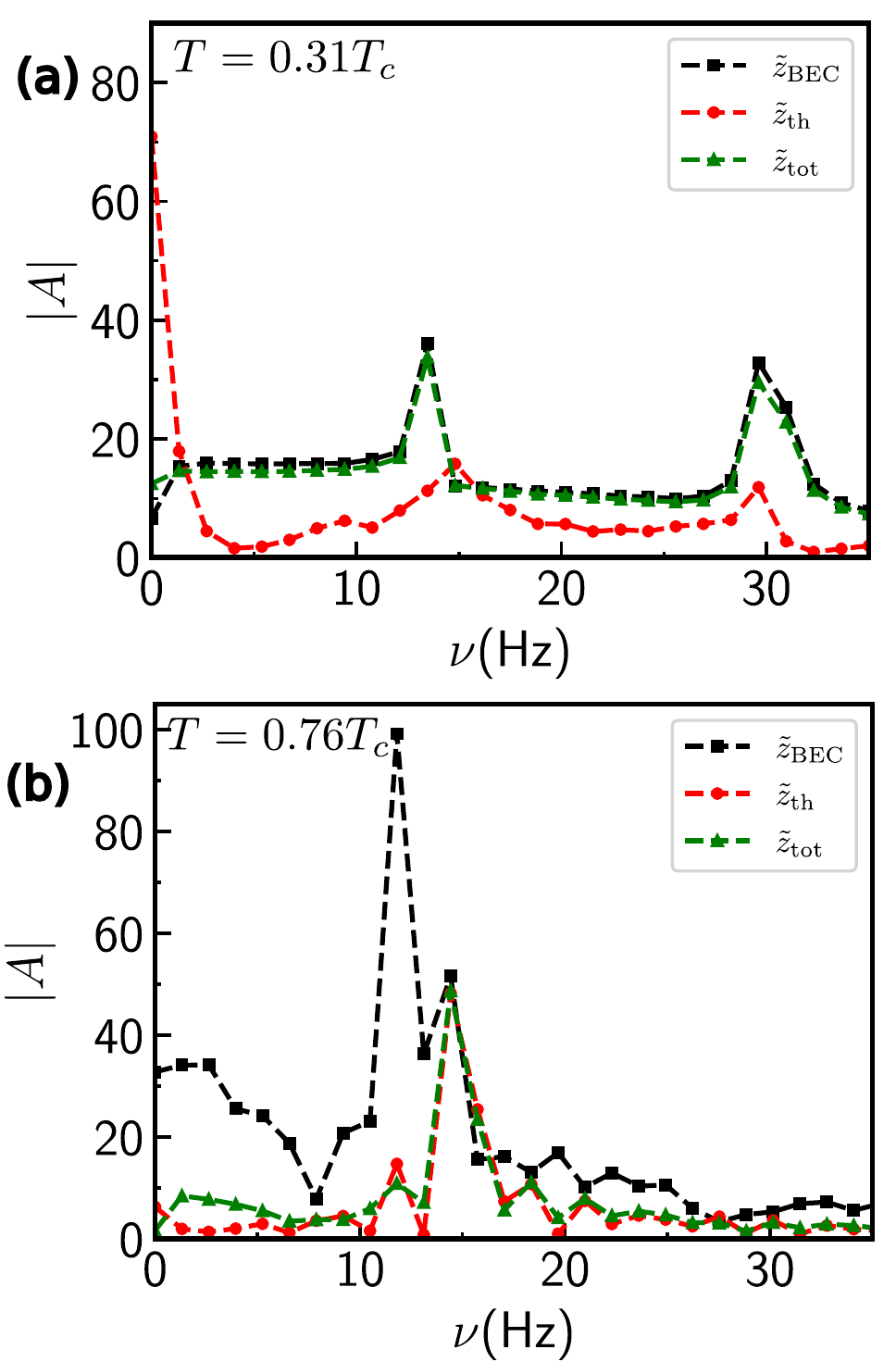}
  \caption{The Discrete Fourier transform (DFT) of the $z_\mathrm{\rm BEC}$(t) (black line), $z_\mathrm{\rm th}$(t) (red line) and $z_\mathrm{\rm tot}$(t) (green line) for $V_0 =104 \hbar \omega _x$ and $z ^\mathrm{\rm BEC} _0\simeq 0.11 >z_\mathrm{cr} ^\mathrm{\rm BEC}$ in the frequency domain for $T=$40nK$=0.31T_c$ (a) and $T=$160 nK$= 0.76 T_c$ (b). The $y$ axis is the amplitude of the components of the DFT $\vert \tilde{z}\vert$ while the $x$ axis is the frequency. The DFT resolution is 1.3Hz.} 
  \label{fig:dft_th_diss}
\end{figure} 

The absolute phase difference between the Josephson and $\nu_1$ contributions is now found to be reduced to about $\pi/2$ up until $k_B T / V_0 \sim 1$, a feature apparently also visible in the total population oscillations. In the latter case, the strong driving of the total particle number by the oscillating thermal cloud leads to approximately equal amplitudes for Josephson plasma and $\nu_2$ contributions to the total population at $T = 0.59T_c$. Moreover, for even larger $T$ the frequency $\nu _2 ^{\rm tot}$ (Fig.~~\ref{fig:nu_diss_bec_tot}(b)(ii)), originating from the effect of the thermal cloud, becomes the dominant total imbalance component.

\begin{figure*}[t!]
\centering
  \includegraphics[width=.99\textwidth]{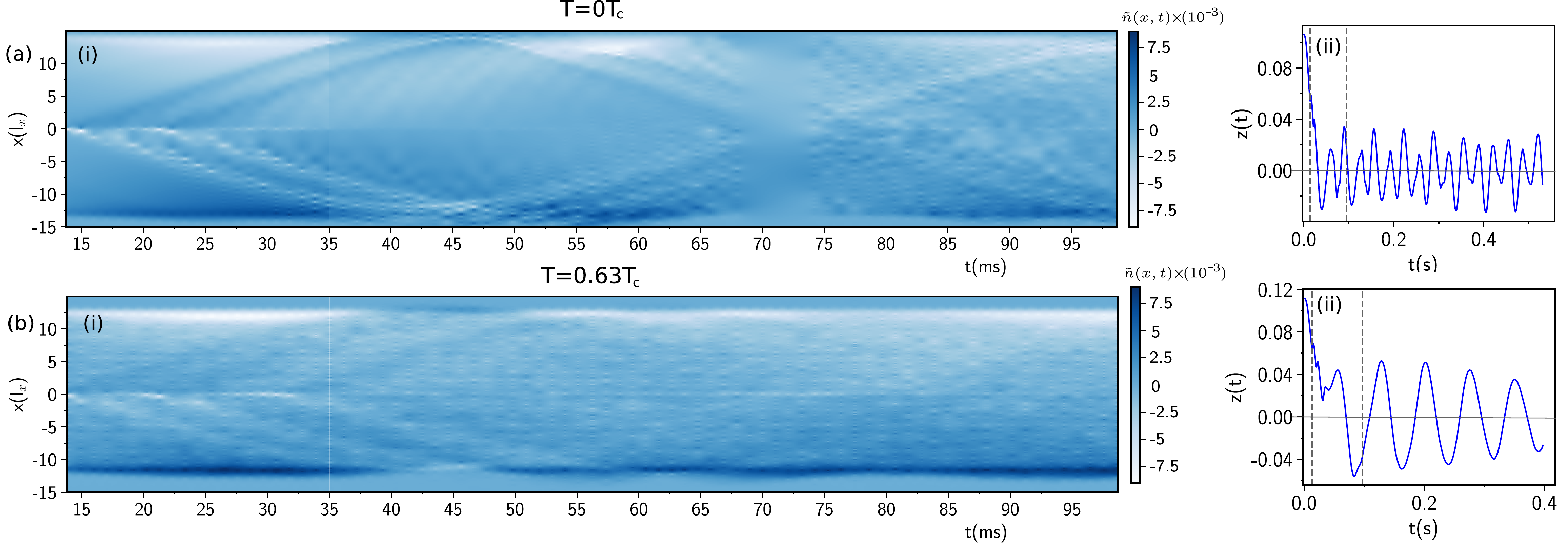}
  \caption{Carpet plots of renormalised density $\tilde{n}(x,t)$ for $z_0 ^{\rm BEC} = 0.11$ (i) at $T=0$ (a) and $T=0.63T_c$ (b) and the corresponding condensate imbalance time evolution (ii) for $V_0=104 \hbar \omega _x$. The vertical grey  dashed lines in (ii) indicate the time interval of the carpet plots.} 
  \label{fig:carpetplot}
\end{figure*}

Looking into more detail at the 3 emerging frequencies, and the damping of the corresponding modes in Fig.~\ref{fig:nu_diss_bec} we note again the similar $\nu_J$ and $\nu_1$ frequency dependence on $T/T_c$ as found earlier in the Josephson plasma  regime (Fig.~\ref{fig:nu_main_bec_jos}); their values decrease with increasing $T$ while the corresponding damping increases with $T$, with $\nu _1$ being damped faster. Meanwhile, $\nu _2$ frequency increases by tending to the $x$-axis trap frequency, while the corresponding damping decreases.

Figure~\ref{fig:dft_th_diss} shows the DFT spectrum of the condensate, the total and the thermal imbalance for a low $T=0.31T_c$ (a) and a high $T=0.76T_c$ (b) temperature. At low $T$ the thermal imbalance spectrum has the same components as the condensate one and the total imbalance spectrum is the same as the condensate one. At high $T$ instead the $\nu_1$ frequency disappears from the spectrum and $\nu _2 \approx 15$Hz appears. It originates from the dominant thermal imbalance dynamics, and manifests itself in both the condensate and total imbalance spectrum. Moreover, at such high $T$ the total imbalance spectrum is close to the thermal one instead. 


Figure~\ref{fig:carpetplot}(i) shows `carpet plots' of the renormalised condensate density $\tilde{n}$ along
the $x$-direction at (a) $T=0$,  and (b) $T=0.63T_c$, with the corresponding population imbalance time evolution shown in Figure~\ref{fig:carpetplot}(ii). In subplots (i), the density $\tilde{n}$ is evaluated by subtracting from the instantaneous density along the $x$ axis, its equilibrium value. In both cases $V_0 \simeq \mu(T=0)$ and thus the vortex rings disappears within the barrier and only the resulting sound waves propagates. Moreover, at relatively long time evolution, sound waves are attenuated due to the presence of the thermal cloud. Thus, the disappearance of $\nu _1$ from the condensate imbalance spectrum coincides with the total damping of sound modes and this confirms the relation between $\nu _1$ and sound waves. 
%
We note that the  temperature $T$ at which sound waves are damped due to the thermal particles depends on the value of $V_0$ and thus such an effect could also occur at smaller $T$ for lower values of $V_0$, such that the thermal energy exceeds the barrier amplitude \footnote[20]{For example, for $V_0/\mu=0.6$ the sound mode is damped already at $T=60$nK}.


%

Our analysis so far has focussed on the role of temperature 
in a system of fixed {\em condensate} particle number -- and thus fixed condensate chemical potential $\mu$ -- which amounts to a
 variable total particle number.
Next, we consider the role of temperature at fixed {\em total} particle number.

\section{Dynamical regimes at Fixed Total Number  \label{sec4}}



When fixing instead the {\em total} number in the system, the effect of temperature is to decrease the condensate number with increasing temperature, due to the increasing presence of particles in the thermal cloud.  This in turn implies that the chemical potential 
becomes temperature-dependent, $\mu(T)$, with a decreasing condensate particle number corresponding to a smaller $\mu(T)$ and smaller spatial extent, both of which significantly affect the system dynamics (see also  Appendix~\ref{app_1} for details).

\begin{figure*}[!htbp]
\centering
  \includegraphics[width=.7\textwidth]{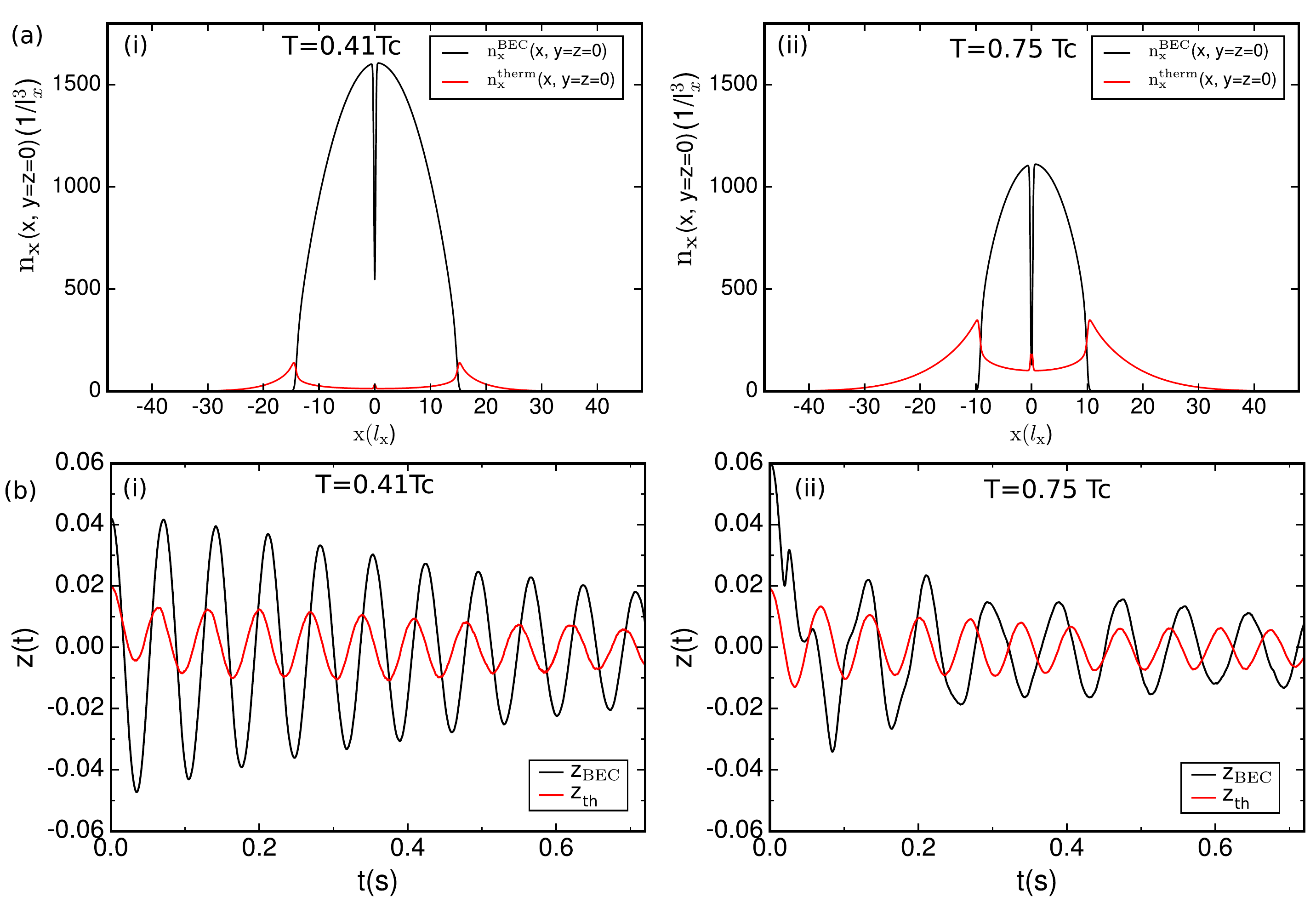}
  \caption{(a) The condensate and the thermal cloud density profiles along the $x$ axis ($y=z=0$) at two different temperatures $T=0.41T_c$ (i) and $T=0.75T_c$ (ii).  (b) The time evolution of the condensate and the thermal cloud  imbalance for T $= 0.41 T_c$ (i) and $T=0.75 T_c$ (ii). The total particle number is kept fixed at $N_\mathrm{\rm tot}=106000$. These data are also for fixed  barrier height $V_0= 104 \hbar \omega _x$ and  barrier shift $\epsilon$ along the $x$ axis.}
  \label{fig:equi_imb_T65_T120_fix_Ntot}
\end{figure*}

This is reflected by the density slices along the main axis (in the $y=z=0$ transversal plane) for both the condensate and the thermal cloud at two different temperatures shown in Fig.~ \ref{fig:equi_imb_T65_T120_fix_Ntot} (a) for fixed $N_ {\rm tot} =106 000$. While the condensate density maxima  decreases with $T$, the thermal cloud maxima instead (at the edges of the condensate),  increases.  Due to the presence of more thermal particles at the edges of the condensate and due to the repulsive interaction between the condensate and thermal particles, the condensate density extension along $x$ axis decreases. Moreover the thermal cloud density at the barrier position increases causing a type of  repulsive potential at the center for the condensate particle and as a consequence the condensate density at $x=0$ decreases.

The tilted linear potential $-\epsilon x$ which is added to the double-well potential at equilibrium in order to generate an initial population imbalance is initially taken to be the same at different $T$. 
Fig.~\ref{fig:equi_imb_T65_T120_fix_Ntot}(b)  shows the corresponding condensate and thermal imbalance temporal evolution for (i) $T = 0.41 T_c$, and (ii) $T=0.75 T_c$. 

Fixing the  {\em total} number of particles while increasing the temperature has two important effects: firstly, the smaller BEC number implies that for the given imposed linear potential, the relative condensate population imbalance increases. Secondly, as $\mu(T)$ decreases with increasing $T$, the ratio $V_0/\mu(T)$  increases (for fixed $V_0$), which is  known   to decrease the value of the critical population imbalance (with all other parameters fixed) marking the transition between Josephson and vortex-induced dissipative regimes (as shown in Appendix C), even at the Gross-Pitaevskii level. 
Thus, for a given external linear potential, increasing temperature at fixed total  number can actually lead to a change in the dynamical regime of the condensate. This is clearly demonstrated in Fig.~\ref{fig:equi_imb_T65_T120_fix_Ntot}(b) showing the
 condensate population imbalance at $T=0.41T_c$ (i) and at $T=0.75 T_c$ (ii): this clearly reveals both that $z_{BEC}^0(T=0.41T_c) < z_{BEC}^0(T=0.75T_c)$ and most significantly, that the condensate population imbalance dynamics transition from the  Josephson regime at (i) $T= 0.41 T_c$ to the vortex-induced dissipative regime at (ii) $T= 0.75 T_c$.
 
\begin{figure}[!htbp]
\centering
 \includegraphics[width=.7\columnwidth]{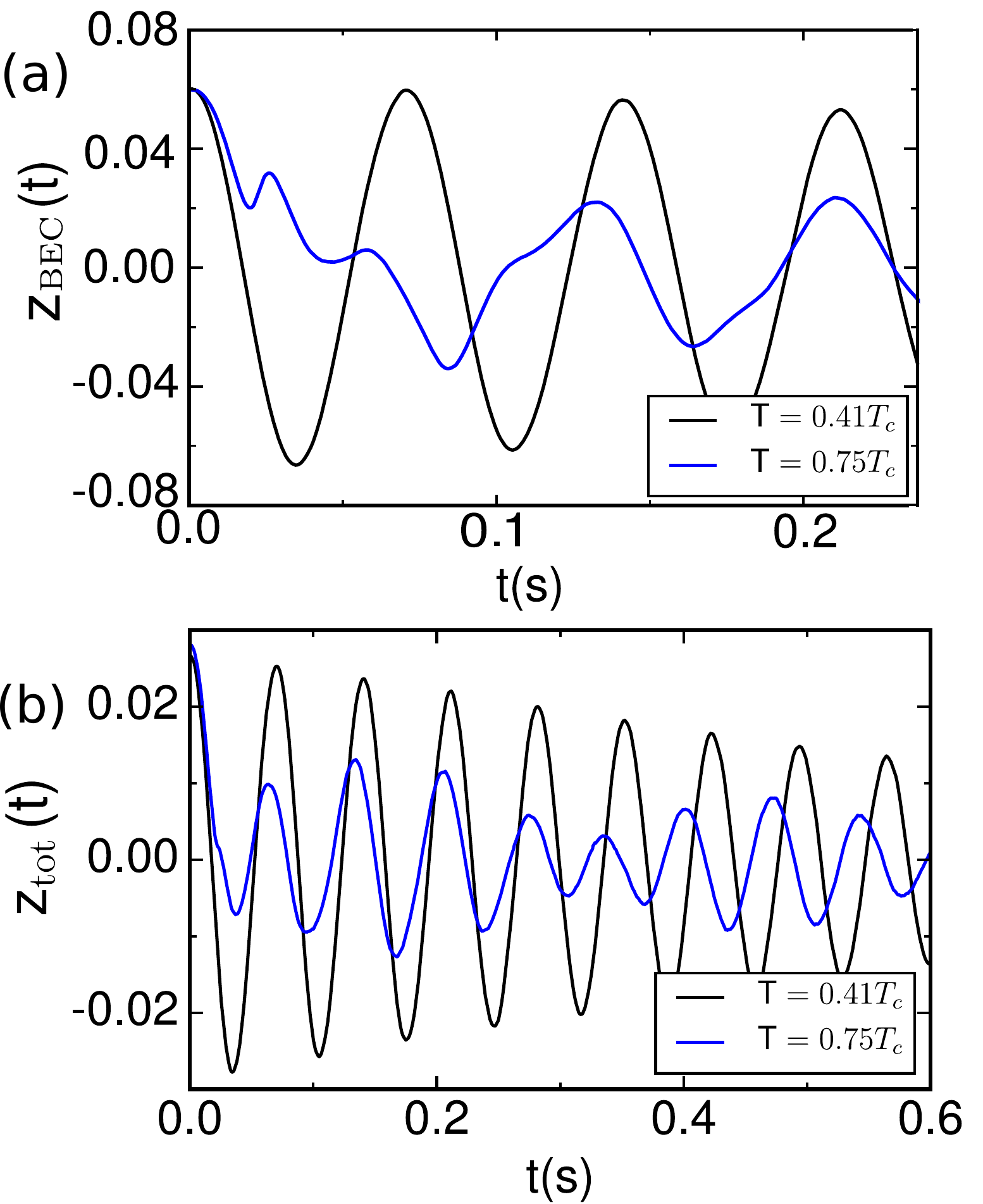}
  \caption{The time evolution of the condensate imbalance (a) and of the total population imbalance (b) in the case when $z_{\rm BEC} ^0$ and $z_{\rm tot} ^0$ are fixed respectively at two different temperatures considered $T= 0.41 T_c$ (black line) and $T= 0.75 T_c$ (blue line). The barrier height is kept fixed at $V_0= 104 \hbar \omega _x$ and fixed total number $N_{\rm tot} =106000$. }
  \label{fig:fig12}
\end{figure}

Related questions of anticipated experimental relevance include the role of particle number, and whether one should be looking at condensate, or total, fractional population imbalance. 
For completeness, we also investigate the difference in the system behaviour between the cases of fixing the condensate, or the total initial population imbalance.
These are shown respectively in Fig.~\ref{fig:fig12}(a) and (b).
It demonstrates clearly that the transition across Josephson and dissipative regimes also occurs with changing temperature in the limit of fixed initial condensate or total population imbalance. The reason for that is that at two different temperatures the ratio  $V_0/\mu$, which defines the system's dynamical regime, is different due to different condensate number.

%

Moreover, we note that, as shown in Appendix C, the critical imbalance at which the system change the dynamical regimes (defined from the condensate imbalance  early-time dynamics) is already determined at the Gross-Pitaevskii level, with the thermal cloud having no significant effect at initial condensate imbalance evolution but strongly affecting its consequent dynamics (long time evolution) as shown in the previous sections.
\section{Discussion \label{sec5a} }

In this paper, we have analyzed for experimentally-relevant conditions the role of thermal dissipation on the superfluid oscillations in the two  dynamical regimes found in an elongated three-dimensional Josephson junction for barrier height close to the chemical potential and barrier width $w/\xi \simeq 4$, namely the Josephson `plasma'  and the vortex-induced dissipative regime. 
The presence of the thermal cloud leads to relative motion between the condensate and the thermal cloud and thus induces dissipative dynamics.
In the Josephson plasma regime, this takes the form of a gradual damping of the condensate oscillations, with an associated decrease in the plasma frequency -- with both effects becoming more pronounced with increasing temperature.
While such a gradual decay mechanism is also at play in the dissipative regime, the short-time evolution in the latter regime exhibits much more drastic dynamics through the
generation (and subsequent dynamics) of vortex rings and associated sound waves, which give a  resistance to the junction that remains finite even at $T=0$ (\cite{Xhani20,Ic2}),
with the thermal cloud having only a comparably small effect in the early-stage dynamics.
Thus, to better characterize the role of the thermal cloud on atomic population dynamics, this work has focussed on the analysis of the long-time evolution of the superfluid dynamics.

Our analysis has revealed the emergence of three dominant frequencies across both probed regimes, with the relative importance of different modes depending on both dynamical regime and temperature.

Firstly, as expected, our study has revealed a Josephson plasma frequency, $\nu_J$, whose value was found to lie slightly below the axial trap frequency $\nu_x$.
While such frequency dominates the low-temperature dynamical behaviour in the Josephson regime, it was found (for the probed experimental parameters) to be subdominant in the dissipative regime. The damping of this plasma oscillation was found to increase super-linearly with temperature, with an associated decrease in $\nu_J$ of up to $\sim18\%$ across both Josephson and dissipative regimes.

At low temperatures $T \lesssim V_0/k_B$ the thermal cloud is relatively small, and is thus primarily driven by the condensate, with a characteristic phase lag.
In this limit, we found an additional frequency which we labelled as $\nu_1$. 
For our chosen parameters, this was found to be $\nu_1 \sim (2.1-2.2) \nu_J \sim 2 \nu_x$.
 The presence of a second frequency at twice the Josephson frequency is expected to be found in the current-phase relation for barrier heights exceeding, but close to, the chemical potential, as already discussed in Refs.~\cite{Ic2,zaccanti_19,Xhani20,singh20}.
 %
 The importance of this mode (which can be interpreted as a phonon-like mode) depends on the dynamical regime.
 In the Josephson plasma regime, this component was found to only play a small, secondary role in the condensate dynamics, leading to a component oscillating with a relative phase of $-\pi$ compared to the plasma oscillation component.
Based on these considerations and knowing that the condensate current is found from the time derivative of the condensate imbalance, the frequency $\nu _1$ is likely to be associated with the non-dissipative  second-order term in the current-phase relation `$\sin (2\Delta \phi)$' originating from the tunneling between condensate and non-condensate states, which at $T=0$ are represented by phonon modes.

Interestingly, in the vortex-induced dissipative regime, the condensate dynamics could be `separated' into short and long time evolution; the former one includes the time interval from the initial time until the time the vortex ring generation ceases (during which the condensate imbalance decays in time) while the later one includes the subsequent dynamics (during which the condensate imbalance oscillates around a zero mean value). 
The short-time evolution defines the maximum superfluid current flowing into the junction and, 
for our geometry, this was previously shown \cite{Xhani20} to depend on both the terms coming from the condensate-to-condensate state tunneling and from condensate-to-noncondensate tunneling;
thus, the arising current-phase relation is sinusoidal of the form $I=I_c \sin (\Delta \phi)-J_1 \sin (2\Delta \phi)$. 
The present work instead focused on the long time evolution, during which the condensate imbalance oscillates around zero value and the current-phase relation is linear. 
As $\nu _1 \approx 2 \nu_J$ and it has an initial relative phase of $\pi/2$ with respect to $\nu _J$, this could lead to the presence of a dissipative component of the form `$\cos (2\Delta \phi)$' in the current-phase relation (due to the presence of a finite chemical potential difference $\Delta\mu $ in our system \cite{Ic2}). Due to the current and phase fluctuations caused by the presence of sound waves and (at finite $T$) also by the presence of thermal cloud, it  is very difficult to extract the importance of such term directly from the current-phase relation.
In fact, in this regime (and for the probed parameters), we found $\nu_1$ to dominate the condensate dynamics across the junction.
This is presumably a direct consequence of the noticeable excitation and propagation of sound waves in the vortex-induced dissipative regime.

As $T \gtrsim V_0/k_B$, the pronounced damping of the sound waves renders the $\nu_1$ frequency practically irrelevant for the condensate imbalance dynamics, thus giving rise to the second frequency $\nu_2 \sim \nu_x$, consistent with dipole oscillations of the incoherent thermal cloud in the underlying harmonic trap, when the thermal cloud acquires sufficient energy to overcome the Gaussian barrier. In fact, in this case, the thermal cloud begins to drive the condensate dynamics, in stark contract to the low-temperature dynamics when the condensate is driving the thermal cloud.

%
Although the relevant frequencies of condensate oscillations beyond the plasma one, and the temperature dependence of their damping and relative phase difference is similar across the Josephson plasma and phase-slip-induced dissipative regimes, we noted a significant difference in their relative importance at low temperatures $T \lesssim k_B T$. In particular we found the $\nu_1$ frequency component to dominate the dissipative regime condensate dynamics at sufficiently low temperatures $T \lesssim V_0/k_B$ -- presumably due to the abundance of sound-wave excitations during the phase slip and subsequent dynamics.
%

Notwithstanding the above comments, condensate dynamics on the high temperature end are in both cases primarily dominated by plasma oscillations (with a significantly reduced frequency), even when the fraction of condensed particles is on the order to $10\%$. Once the dynamical thermal cloud is however included, the total particle evolution exhibits a combination of self-driven plasma oscillations and thermal cloud oscillations in the trap, which lead to noticeable beating in both condensate and total population imbalances, an effect which is within current experimentally reach.

In order to keep a fixed condensate size and ratio of $V_0/\mu(T=0)$, our analysis was conducted for fixed condensate number, thus implying a variable critical temperature for condensation.
Our analysis was further extended to the case of
a fixed total particle number, and we showed that having different condensate numbers at different $T$ can make the condensate or total population imbalance to be in a different dynamical regime  due to different  $N_{\rm BEC}$ and $V_0/\mu$ values. 


\section{Conclusions  \label{sec5}}

In brief, we characterised through state-of-the-art numerical simulations the temperature dependence and damping of dominant dynamical excitation modes of a finite-temperature superfluid across a thin Josephson junction, which supports a transition from plasma to dissipative phase-slip-induced regimes.
Beyond the characterisation of the plasma mode, we identified a further relevant low-temperature and high-temperature mode, distinguished by the ratio $k_B T/V_0$ of thermal to barrier energy; the thermal dynamics were shown to lead to damping of the condensate motion, with a new regime identified in which the (dominant) thermal cloud has enough energy to overcome the axial barrier and thus begins to drive the condensate out of phase, leading to beating in the condensate and/or total population dynamics.
The additional frequency emerging in the low temperature limit
in the vortex-induced dissipative  (or Josephson `plasma') regime
was attributed to  a second-order dissipative (non-dissipative) term  in the superfluid current, which derives from the tunneling between condensate to non-condensate states.
Our findings, based on an established self-consistently coupled kinetic model, are within current experimental reach in ultracold superfluid junctions.

Data supporting this publication are openly available under
an “Open Data Commons Open Database License”\footnote{Link to be inserted prior to publication}.

\section*{Acknowledgement}
We acknowledge discussions with Kean Loon Lee, I-Kang (Gary) Liu, Giacomo Roati, Francesco Scazza,   and Matteo Zaccanti,
and financial support from the QuantERA project NAQUAS
(EPSRC EP/R043434/1), 
and Qombs Project [FET Flagship on Quantum Technologies grant n. 820419].

\appendix

\section{Modelling Scheme Details \label{app_1}}

Below, we give a brief summary of the kinetic model used in our work.

At finite temperature, the system wavefunction is written as the sum of a condensate wavefunction and a thermal cloud. The condensate wavefunction evolution is found by solving the generalized Gross-Pitaevskii equation
which  accounts for the thermal cloud mean field potential, $2g n_\mathrm{th}$ \cite{ZNG2}:
\begin{equation}
\label{gped}
i \hbar \frac{\partial \psi}{\partial t}=\left[ - \frac{\hbar^2 \nabla ^2}{2M}+V_\mathrm{ext}+g(|\psi|^2+2n_\mathrm{th})\right] \psi \;,
\end{equation}
with other symbols defined in the main text.

To initiate the dynamics, we first obtain the equilibrium solution in a static potential including the harmonic trap, Gaussian barrier, and linear potential. 
The equilibrium condensate wavefunction $\psi _0$ is obtained self-consistently via 
\begin{equation}\label{gped2}
\mu(T) \psi _0=\left( -\frac{\hbar ^2}{2M} \nabla ^2  + V_\mathrm{ext} + g (|\psi _0 |^2 + 2 n^0 _\mathrm{th})\right)  \psi_0
\end{equation}
where $n^0 _\mathrm{th}$ and $n^0 _\mathrm{BEC}=|\psi _0 |^2$ are the equilibrium thermal and condensate density respectively,  while $\mu(T)$ is the temperature-dependent system chemical potential accounting for the thermal cloud equilibrium mean field potential. 
 $V_{ext}(\bold{r})$ is the trapping potential defined by Eq.~(\ref{Vtrap}).
As described in \cite{ZNG3,ZNG6,nick_book} the initial thermal cloud density ansatz is based on a simple Gaussian for the required temperature.
We then iterate those 2 equations self-consistently, until arriving at an equilibrium solution with the desired condensate, or total, particle number at each specified temperature.

\begin{figure}[t!]
\centering
  \includegraphics[width=\columnwidth]{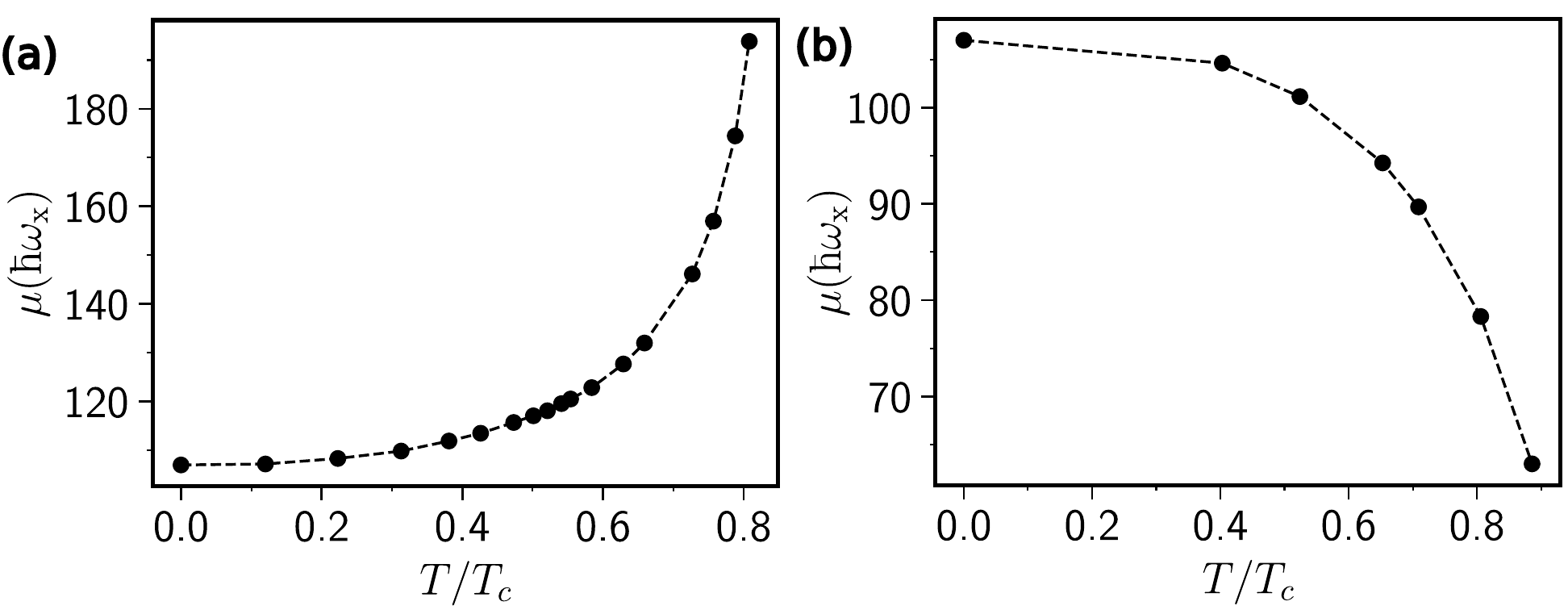}
  \caption{The chemical potential as a function of the temperature $T/T_c$ for fixed condensate number $N_{BEC}\simeq 50400$ (a) and fixed total number $N_{tot}= 50400$ (b). Both plots are obtained for the same  barrier height value $V_0=104 \hbar \omega _x$.}
  \label{fig:fracT_muT}
\end{figure}

The presence of the thermal cloud modifies the value of the chemical potential $\mu$ of the system which takes into account also the mean field potential of the thermal cloud. 
The main analysis in this paper is conducted at fixed condensate particle number $N_{\rm BEC} \simeq  5.04  \times 10^4$: for such parameters, the dependence of $\mu$ on the scaled temperature is shown in Fig.~\ref{fig:fracT_muT}(a) which increases with $T$ due to the increasing thermal fraction. 
In the opposite case of fixed total number, the temperature dependence of the chemical potential is shown in Fig.~\ref{fig:fracT_muT}(b), revealing a decreasing dependence on temperature (due to the decreasing condensate number).
%

Changing temperature while keeping $N_{\rm BEC}$ fixed also changes $N_{\rm tot}$, and hence the corresponding non-interacting critical temperature $T_c=T_c(N)$.
In our analysis, we specifically probe the temperature regime $T/T_c \lesssim 0.8$, for which the condensate fraction $N_{\rm BEC} / N_{\rm tot} \in [0.1,\,1]$.

In the Hartee-Fock limit, and in the presence of an external potential,
the energy of a particle becomes 
\begin{equation}
\epsilon(\bold{r},t)=\frac{\bold{p}^2}{2m}+V_\mathrm{ext}(\bold{r})+ 2g (n_\mathrm{BEC}(\bold{r},t)+n_\mathrm{th}(\bold{r},t))\;.
\end{equation}  
Thus the thermal particle feel a generalized effective potential:
\begin{equation}
V_\mathrm{eff} ^{\mathrm{th}}(\bold{r},t)= V_\mathrm{ext}(\bold{r})+2g(n_\mathrm{BEC}(\bold{r},t)+n_\mathrm{th}(\bold{r},t))
\end{equation}
  while the condensate atoms feels an effective potential given by :
\begin{equation}
V_\mathrm{eff} ^{\mathrm{BEC}}(\bold{r},t)= V_\mathrm{ext}+g(n_\mathrm{BEC}(\bold{r},t)+2n_\mathrm{th}(\bold{r},t))
\end{equation}

To seed the oscillatory dynamics, at $t=0$, the linear potential is instantaneously removed, and the subsequent dynamics of all components are analyzed in detail.  
To account for thermal cloud dynamics, we solve this equation self-consistently with a collisionless Boltzmann equation for the thermal molecule phase-space distribution, $f$, obeying: 
\begin{equation}
\label{bolt}
\frac{\partial f}{\partial t}+ \frac{\bold{p}}{M} \cdot \nabla_{\bold{r}} f-\nabla_{\bold{r}} V_\mathrm{eff} ^{\mathrm{th}} \cdot \nabla_{\bold{p}}f=0
\end{equation}
where  the thermal cloud density is defined by 
\begin{equation}
n_{th}=\frac{1}{(2\pi\hbar)^3} \int d\bold{p} ~ f(\bold{p},\bold{r},t) \;.
\end{equation}
Our model corresponds to the collisionless limit of the ``Zaremba-Nikuni-Griffin'' (ZNG) kinetic theory. Refs.~\cite{vortexT1, ZNG2, nick_book, vortexT2, vortexT3,ZNG4,ZNG6} shows that this theoretical model successfully describes the collective modes, 
 vortex dynamics and evaporative cooling.

 \section{Equilibrium state \label{app_2}}
\begin{figure}[t!]
\begin{center}
\includegraphics[width= 0.8\columnwidth]{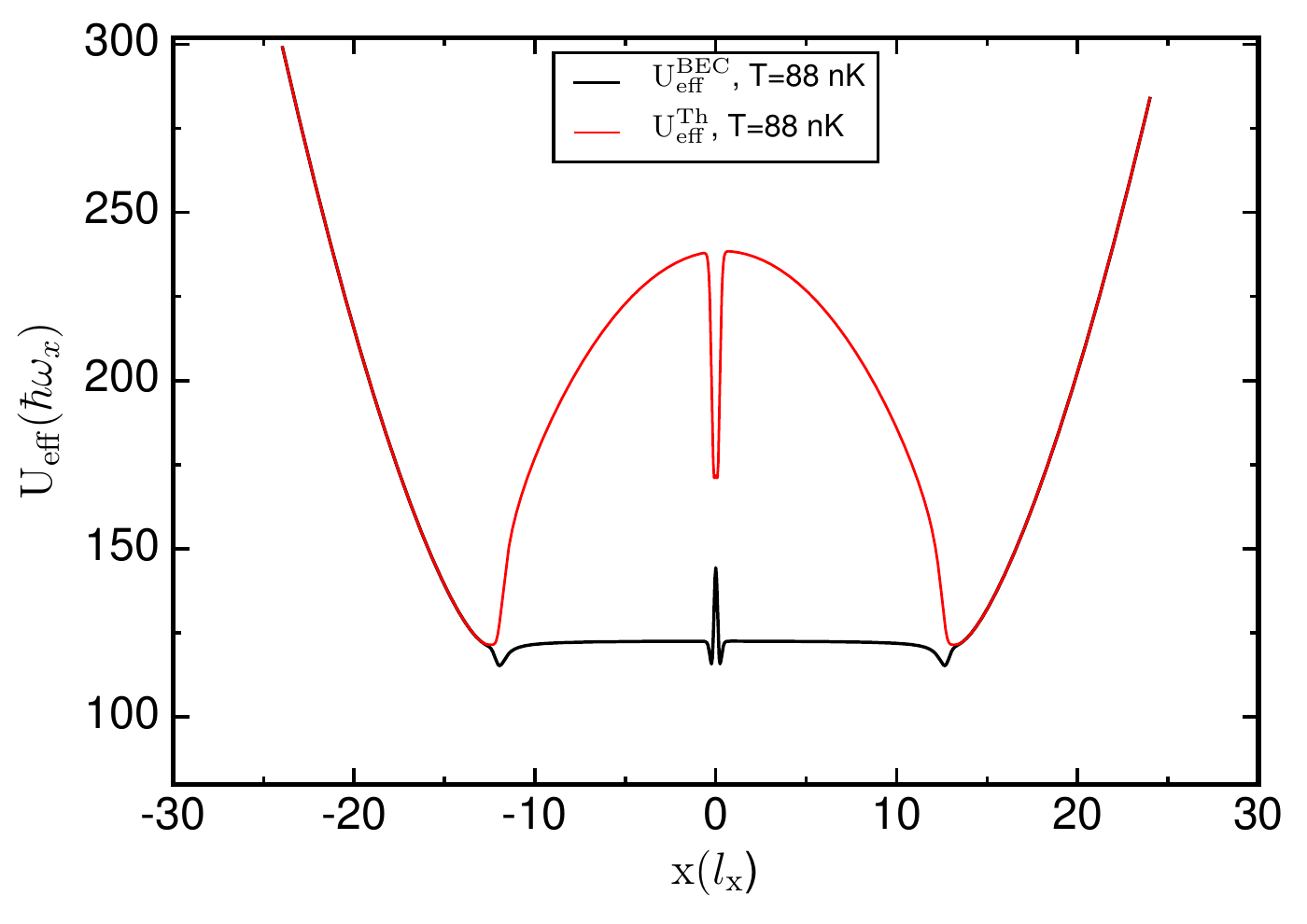}
\vspace{-0.3cm}
\caption{
The equilibrium effective potential felt by the condensate (black curve) and by the thermal cloud (red curve) at $T=$88nK $= 0.58 T_c$ for $V_0=104 \hbar \omega _x$.}  
\label{fig:U_eff}
\end{center}
\end{figure}

The equilibrium effective potential profile felt by the condensate and the thermal particle for a double-well potential are shown in Fig.~\ref{fig:U_eff}, respectively by black and red lines. 
 Due to repulsive interaction between particles (i.e. g$>$0) the thermal cloud feels a larger potential where the condensate density is larger, which means that the thermal particles would have lower density. This explains why the thermal cloud density has local maxima at the barrier position (where the condensate density is minimum) and at the edges of the condensate.
 

\begin{figure}[t!]
\centering
  \includegraphics[width=.6\linewidth]{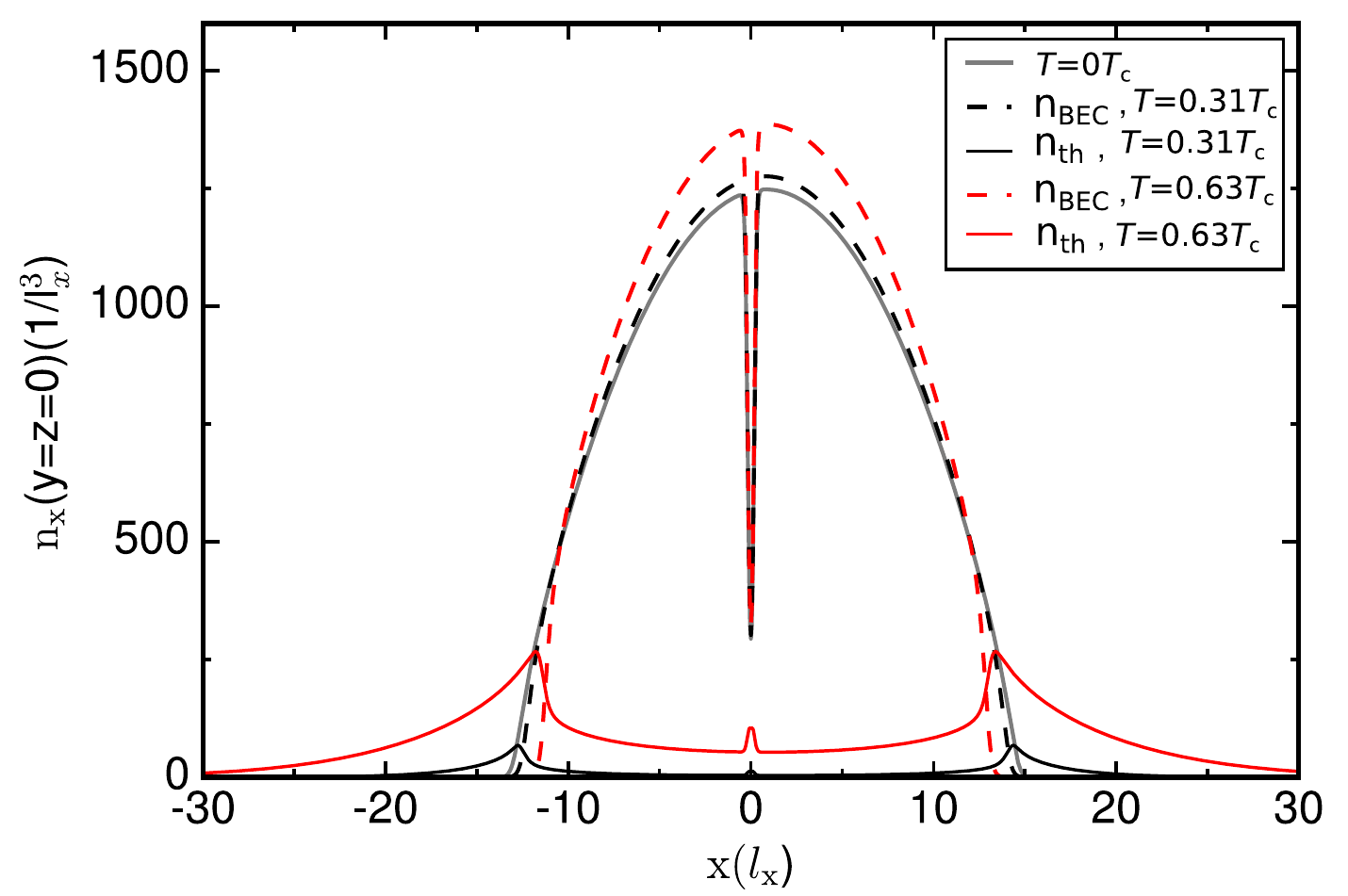}
  \caption{The condensate density profile along the $x$ direction ($y=z=0$)  for three  different temperatures: $T=0$ (grey line), $T=$40nK $=0.31 T_c$ (black dashed line) and $T=$100 nK $= 0.63 T_c$ (red dashed line) and the  thermal cloud density shown as black solid line for $T=0.31 T_c$ and as red solid line for $T= 0.63 T_c$.  The barrier height is fixed at $V_0=104 \hbar \omega _x$, $w/\xi \simeq 4$ and $N_{\rm BEC}\simeq 5.04 \times 10^4$.}
  \label{fig:den_1D_T50_88}
\end{figure}

For comparison, Fig.~\ref{fig:den_1D_T50_88} shows  the equilibrium condensate and thermal density along the $x$ axis for $y=z=0$ for three different temperatures: $T=0$,  $T = 0.31 T_c$ and $T= 0.63 T_c$. The numerical data are obtained for $V_0=104 \hbar \omega _x \simeq \mu (T=0)$  and in the presence of an initial imbalance. 
In all three case, the condensate number is kept fixed at $N_{\rm BEC}\simeq 5.04 \times 10^4$. 
We note that the presence of the thermal cloud at the edges of the condensate makes the condensate density to have a slightly smaller $x$ axis extension with respect to the condensate at $T=0$ where the thermal cloud is not present and thus a larger maximum density in order to keep the condensate number fixed. This effect is stronger at a higher temperature $T=100 \rm{nK} = 0.63 T_c$. 


 \section{The critical imbalance\label{app3}}
\begin{figure}[t!]
\begin{center}
\includegraphics[width= \columnwidth]{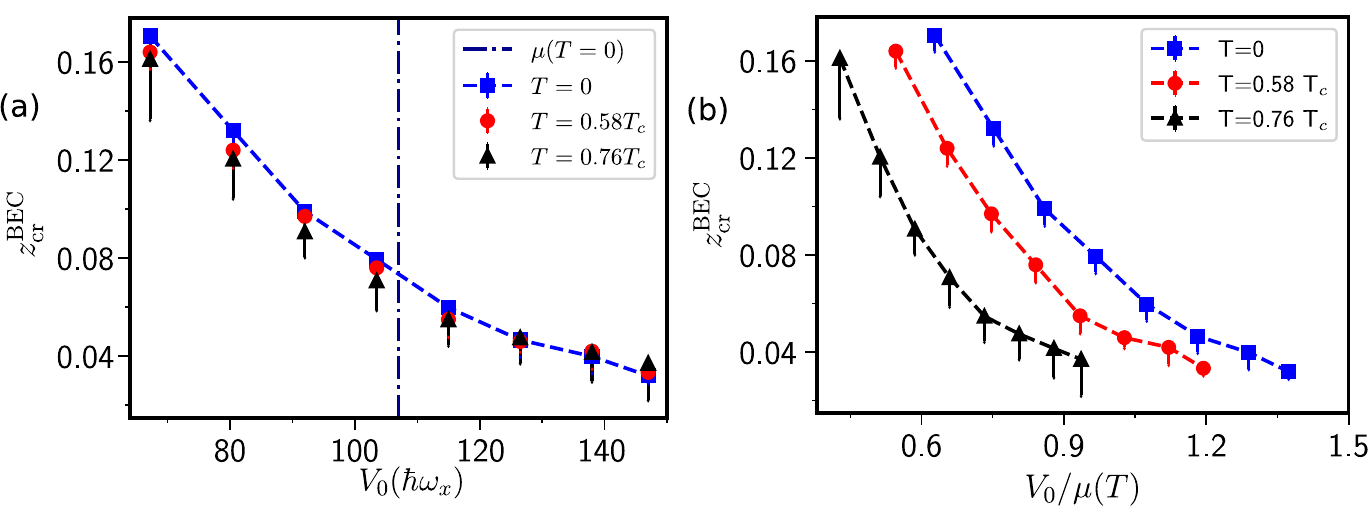}
\vspace{-0.3cm}
\caption{The critical condensate initial imbalance at $T=0$,  $T=0.58T_c$ and $T=$160nK=$0.76T_c$ as a function of $V_0$ (a) and $V_0/\mu(T)$ (b). These data are for fixed condensate number when comparing $T=0$ and finite $T$ results. The dashed line in (a) indicate the chemical potential at $T=0$.}  
\label{fig:zcr_muT}
\end{center}
\end{figure}

 Based on the short time evolution of the condensate population imbalance a critical value of initial imbalance is found $z_{\rm cr} ^{\rm BEC}$, which is defined as the value of initial imbalance where only one phase-slippage occurs and thus only one vortex ring is generated. Figure~\ref{fig:zcr_muT} (a) shows  the profile of the critical imbalance as a function of the barrier height value, and  (b) the corresponding plot with the barrier height scaled to the chemical potential at finite $T$. In these plots, the condensate number is fixed at $N_{\rm BEC}\simeq 5.04 \times 10^4$, which means $N_{\rm tot }$ varies with $T$. We firstly note that as $T$ increases and for fixed $V_0$ (i.e.~$V_0/\mu(T=0)$) the critical imbalance is not affected by the presence of the thermal cloud. If instead, the $V_0/\mu(T)$ is fixed, the critical condensate imbalance at finite $T$ differs from its value at $T=0$ and this difference is larger for larger values of $T/T_c$.
 In fact, at $T=0.3T_c$ as in our previous studies where the condensate fraction is 90$\%$ and within numerical error bars, $z_{\rm cr} ^{BEC}(T=0.3T_c) \simeq z_{\rm cr} ^{\rm BEC}(T=0)$  for the same $V_0/\mu(T)$. 
 Moreover, as $T$ increases, $z_{\rm cr } ^{\rm BEC}$ shifts more from its corresponding $T=0$ value, thus being smaller for the same value $V_0/\mu(T)$. 
 
 \section{The vortex ring at finite T}
\begin{figure}[h!]
\begin{center}
\includegraphics[width= \columnwidth]{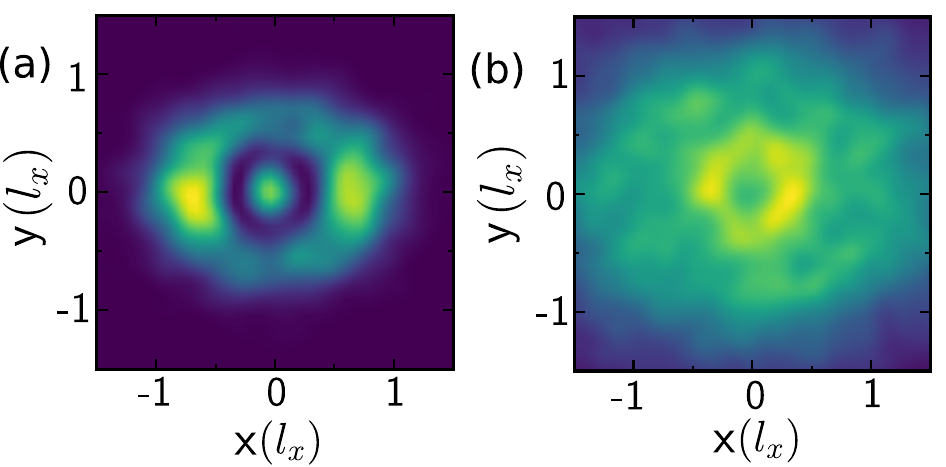}
\vspace{-0.3cm}
\caption{The condensate (a) and thermal cloud density (b) along the $yz$ plane and in the presence of a vortex ring. This data are for $T=0.58T_c$ and for $V_0/\mu(T=0)=0.6$}  
\label{fig:VR}
\end{center}
\end{figure}

 In our 3D geometry, every time the superfluid velocity exceeds a critical value, vortex rings are generated. The vortex ring has a core which is characterized by a vanishing condensate density as visible in Fig.~\ref{fig:VR}(a). At finite temperature, the repulsive interactions between the condensate and thermal cloud densities, can lead the vortex ring core to becomes itself populated by the thermal cloud as shown in Fig.~\ref{fig:VR}(b).

 \section{The role of the thermal cloud on initial condensate imbalance \label{app4}}
As we showed in the main text (Sec.~III C and Fig.~6), for $T \gtrsim 0.58 T_c$ and in the vortex-induced dissipative regime, there is a third `dip' (or kink) in the initial decay of the condensate population imbalance, corresponding to the generation of an additional vortex ring. We also noted that 
\begin{figure*}[!t]
\centering
  \includegraphics[width=.8\linewidth]{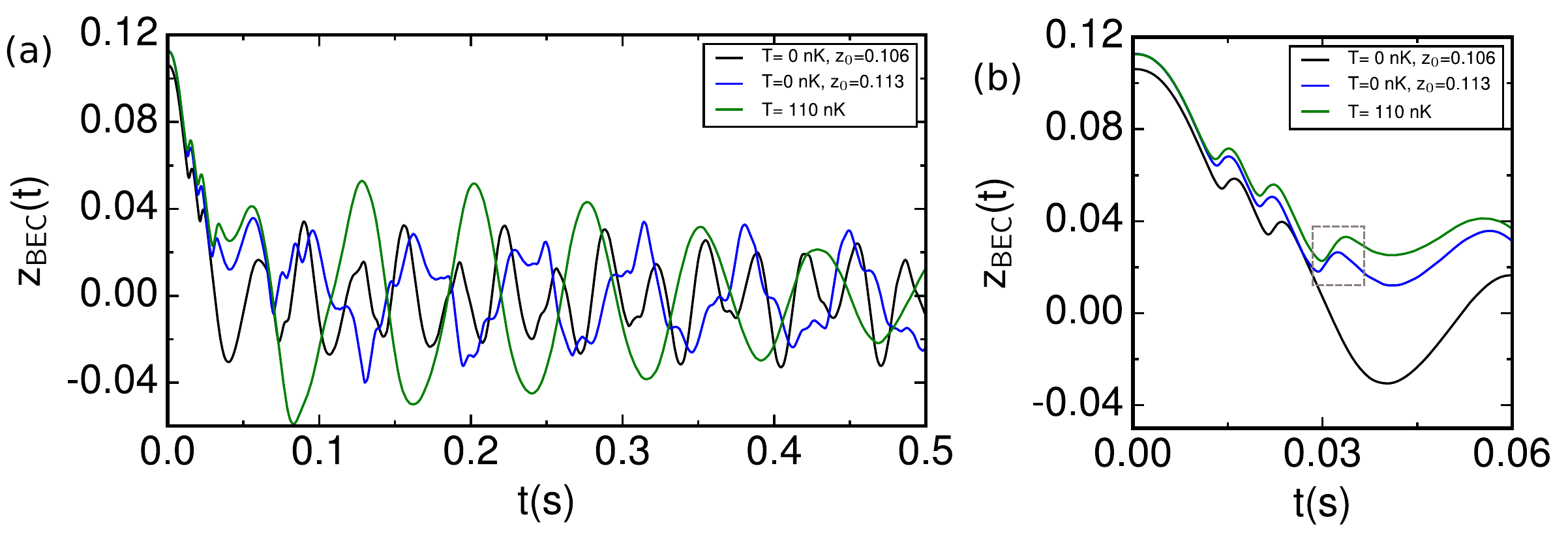}
  \caption{(a) The condensate population imbalance time evolution at $T=0$ and for two slightly different $z_0$ and at $T=110 \rm{nK}=0.66T_c$ with a zoomed-in corresponding profile shown in (b).  The data are for $V_0 = 104 \hbar \omega _x$ and fixed $N_\mathrm{\rm BEC} \simeq 50400$.}
  \label{fig:zc_T0_diff_eps_T110}
\end{figure*}
the concentration of the thermal cloud at the condensate edges and at the barrier induces a small shift in the condensate initial imbalance which slightly increases with temperature. 
 The first question is whether the new shifted $z_{\rm BEC} ^0$ would have been large enough at $T=0$  to cause the transition to a regime when another phase slippage happens or it comes from the thermal induced fluctuations of the phase. The second question is related to the reason why  at $T=0.58T_c$ the $\nu _1 ^{\rm BEC}$ `disappears' from the spectrum of the condensate imbalance when for $T<0.58T_c$ its contribution is large, around $60\%$.  

In order to answer such questions, we consider also the case when the initial condensate imbalance is fixed between the cases of $T=0$ and the selected $T=0.66T_c$, instead of fixing $\epsilon$ of the lineal potential $-\epsilon x$.
Fig.~\ref{fig:zc_T0_diff_eps_T110} shows the condensate population imbalance for $T=0$ and $T=110 \rm{nK}= 0.66 T_c$, for two slightly different initial population imbalances $z_0 = 0.106$ and  $z_0 = 0.113$.
These values of $z_0$ have been chosen such that the first one produces the same barrier shift as $T= 0.66 T_c$, while the second 
one is exactly equal to the initial condensate imbalance $z_{\rm BEC} ^0$ at  $T= 0.66 T_c$.

We observe that at $T=0$, the condensate population imbalance  for $z_0 = 0.113$, which is equal to the initial condensate imbalance at $T=0.66T_c$, presents a third `dip', indicated by the white rectangle in the zoomed-in profile in Fig.~\ref{fig:zc_T0_diff_eps_T110}(b).  Thus this analysis explains the origin of the third generated vortex ring (for $V_0=104 \hbar \omega _x$).
We note that the initial decay of the condensate imbalance is similar for $[T=0,\ z_0 = 0.113]$ and $T=0.66T_c$.  
Furthermore, the temporal profile of the population imbalance at $T=0$, after the initial decay, shows different features for $z_0 = 0.113$ with respect to $z_0 = 0.106$. In particular in the first case the frequency near 30Hz, i.e. the $\nu _1 ^{\rm BEC}$ frequency, is less important than the Josephson `plasma' frequency. As we have shown previously, the opposite happens for $z_0 = 0.106$, where the $\nu _1$ is the  dominant frequency. Thus this answers to our second question. 
Moreover, from Fig.~\ref{fig:zc_T0_diff_eps_T110} we observe also that the   long-time evolution differs between [$T=0$,$z_0=0.113$] and $T=0.66T_c$, i.e. for the same initial condensate imbalance between $T=0$ and $T=0.66T_c$, due to the presence of a significant thermal fraction at this temperature which damps second-order term. 

\section*{References}

\bibliography{klejdja_biblio_SM_AT}
\end{document}